\documentclass[a4paper,onecolumn,oneside,fleqn,10pt]{article}
\usepackage{abstract}
\usepackage{amsmath}
\usepackage{authblk}
\usepackage[english]{babel}
\usepackage{bm}
\usepackage[font={color=blue},figurename=Fig.,labelfont={it}]{caption}
\usepackage{cite}
\usepackage{enumitem}   
\usepackage{fchicago} 
\usepackage{float}
\usepackage[left=2cm,right=2cm,top=2cm,bottom=2cm]{geometry}
\usepackage{graphicx}
\usepackage{hyperref}
\usepackage{indentfirst}
\usepackage[utf8]{inputenc}
\usepackage[switch,columnwise]{lineno}
\usepackage{lipsum} 
\usepackage{pdfpages}
\usepackage{rotating}
\usepackage{setspace} 
\usepackage{subcaption}
\usepackage[varg]{txfonts}
\usepackage{wasysym}
\usepackage{fmtcount} 

\definecolor{blue}{RGB}{0,0,255}
\hypersetup{
    colorlinks=true,                          
    linkcolor=blue, 
    citecolor=blue, 
    urlcolor=blue,
    linktoc=all
}

\newcommand{\via}{\textit{via }}
\newcommand{\etc}{\textit{etc }\ldots }
\newcommand{\eg}{\textit{eg. }}
\newcommand{\cf}{\textit{cf. }}
\newcommand{\ie}{\textit{ie }}

\usepackage{fancyhdr}
\graphicspath{{images/}{figures/pdf/}}

\usepackage{listings}
\usepackage{xcolor}  

\begin{document}

\title{\huge On a planetary forcing of global seismicity \vspace{2cm}}

\author[1]{Dumont Stéphanie}
\author[2]{de Bremond d’Ars Jean}
\author[3]{Boulé Jean-Baptiste}
\author[4]{Courtillot Vincent}
\author[5]{Gèze Marc\footnote{$\dagger$Marc Gèze passed away before the finalization of this manuscript. We dedicate this work to his memory.}}
\author[6]{Gibert Dominique}
\author[7]{Kossobokov Vladimir}
\author[4]{Le Mouël Jean-Louis}
\author[3]{Lopes Fernando}
\author[8]{Neves Maria C.}
\author[9,10]{Silveira Graça}
\author[11]{Petrosino Simona}
\author[12]{Zuddas Pierpaolo\vspace{2cm}}

\affil[1]{\small Instituto Dom Luiz (IDL), Faculdade de Ciências, Universidade de Lisboa, Lisboa, Portugal}
\affil[2]{\small Univ Rennes, CNRS, Géosciences Rennes, UMR 6118, Rennes, France}
\affil[3]{\small Muséum National d’Histoire Naturelle, CNRS UMR7196, INSERM U1154, Paris, France}
\affil[4]{\small Académie des Sciences, Institut de France, Paris, France}
\affil[5]{\small Muséum National d’Histoire Naturelle, CEMIM, Sorbonne Université, Paris, France}
\affil[6]{\small LGL-TPE, Univ Lyon, Univ Lyon 1, ENSL, CNRS, UMR 5276, Villeurbanne, France}
\affil[7]{\small Institute of Earthquake Prediction Theory and Mathematical Geophysics, Russian Academy of Sciences, Moscow, Russia}
\affil[8]{\small FCT, Campus de Gambelas, Universidade do Algarve, Faro, Portugal}
\affil[9]{\small Instituto Dom Luiz (IDL), Faculdade de Ciências, Universidade de Lisboa, Lisboa, Portugal}
\affil[10]{\small Instituto Superior de Engenharia de Lisboa (ISEL), Instituto Politécnico de Lisboa, Lisboa, Portugal}
\affil[11]{\small Instituto Nazionale di Geofisica e Vulcanologia, Sezione di Napoli—Osservatorio Vesuviano, Naples, Italy}
\affil[12]{\small Sorbonne Université, CNRS, METIS, Paris, France}

\date{}
\maketitle

\newpage

\begin{abstract}
	We have explored the temporal variability of the seismicity at global scale over the last 124 years, as well as its potential drivers. To achieve this, we constructed and analyzed an averaged global seismicity curve for earthquakes of magnitude equal or greater than 6 since 1900. Using Singular Spectrum Analysis, we decomposed this curve and compared the extracted pseudo-cycles with two global geophysical parameters associated with Earth’s tides: length-of-day variations and sea-level changes. Our results reveal that these three geophysical phenomena can be be explained with 90\% accuracy, as the sum of up to seven periodic components, largely aligned with planetary ephemerides: 1 year, 3.4 years (Quasi-Biennial Oscillation, QBO), $\sim$11 years, $\sim$14 years, $\sim$19 years (lunar nodal cycle), $\sim$33 years, and $\sim$60 years.  We discuss these results in the framework of Laplace’s theory, with a particular focus on the phase relationships between seismicity, length-of-day variations, and sea-level changes to further  elucidate the underlying physical mechanisms. Finally, integrating observations from seismogenic regions, we propose a trigger mechanism based on solid Earth–hydrosphere interactions, emphasizing the key role of water-rock interactions in modulating earthquake occurrence.
	
\end{abstract}

\section{\label{sec01} Introduction} 
	The question of apparent random nature of earthquakes remains an important and contemporary topic (\eg \shortciteNP{Gardner1974,Heaton1975,Klein1976,Kilston1983,Mazzarella1989,Lopes1990,Kossobokov2006,Metivier2009,Hough2018,Varga2019, Kossobokov2020}). An earthquake is a sudden movement within the Earth’s lithosphere. Earthquake occurrences are not random, but rather haphazard, i.e. lacking any obvious principle of organization. Mathematically, the characteristics of such haphazard systems, apparently chaotic, are nevertheless predictable up to a certain limit and after substantial averaging. In particular, the results of on-going global testing of the earthquake prediction algorithm M8 started in 1992 (\eg \shortciteNP{Healy1992,IsmailZadeh2021}) have provided evidence of predictability for most of the world largest earthquakes, although up to a certain space-time limit of intermediate-term middle-range accuracy.
	
	For seismic events, The notion of "preferred days" or “astronomical forcing” is an ancient concept first found in the writings of Pliny the Elder (77 AD), with a renewed interest due to \shortciteN{Perrey1875}, who regarded lunar forces as the primary driver. Regardless of the presumed nature of this forcing, studies have long tended not merely to minimize but to invalidate the concept of external forcing, such as tidal effects, by consistently relying on fracture and stress/strain mechanics principles (\eg \shortciteNP{Rydelek1992,VereJones1995,Jordan2011}), and therefore on physical considerations, with seldom reference to (geo)chemical processes. 	
	
	Regarding regional and global seismicity, independent studies conducted over time have used spectral analyses to investigate strong, significant or even great earthquakes (\eg \shortciteNP{Liritzis1993,Malyshkov2009,Scafetta2015}). These studies have identified periodicities around $\sim$5 years, $\sim$7 years, $\sim$8 years, $\sim$11 years, $\sim$14 years, $\sim$18 years, $\sim$20 years, and $\sim$40 years. Of course, it is important to bear in mind that spectral analysis is known to be challenging for strictly non-stationary signals (\eg \shortciteNP{Kay1981} or \shortciteNP{Gibert2024}, chapter 11). Comparing these results is further complicated by variations in the datasets used, which often differ in spatial scale (regional vs. global) and magnitude thresholds, making direct comparisons challenging. Nevertheless, these periodicities are to some extent consistent with Laplace commensurabilities, i.e., orbital resonances of celestial bodies (\cf \shortciteNP{Morth1979,Lopes2021}). These periods, which are more or less compatible over a common timescale with those observed for average global volcanism (\cf \shortciteNP{Dumont2022,LeMouel2023}), are particularly consistent with fluctuations in mean sea level (\eg \shortciteNP{Courtillot2022}) since the establishment of the first tide gauge in Brest in 1807 (\cf \shortciteNP{LeMouel2021}). These fluctuations are also known to mirror those of continental groundwater levels (\eg \shortciteNP{Russo2017,Liesch2019,Diodato2024,Fan2024,Neves2016}). The aforementioned tidal periods are associated with the Jovian planets and are an order of magnitude smaller than those of lunisolar tides (\eg \shortciteNP{Lambeck2005,LeMouel2019a}). Nevertheless, they have the advantage of being globally effective over extended time scales ($\geq$ 1 year) across the entire Earth, unlike shorter tidal potentials that manifest in tesseral, zonal, and sectoral distributions on our globe (\eg \shortciteNP{Ray2014,LeMouel2024}), \ie a lunisolar tide does not uniformly affect the same location on Earth within a single day.

	Variations in the global hydrosphere are relatively well characterized, being primarily periodic (\shortciteNP{McMillan2019,Diodato2024,Fan2024}). Conversely, water-rock interactions and their kinetics are much more complex (\eg \shortciteNP{Brantley2008,Traskin2009,Zuddas2010}), depending on water and mineral composition, depth of water tables, their interconnections, and much more. Such complexity inherently precludes any local short-term predictability of seismic activity. Nevertheless, on longer time scales, this water-rock interaction effect is likely to re-emerge either locally, \eg in phenomena related to mineral dissolution (\shortciteNP{Reddy2012,Barberio2017,Zuddas2024,Skelton2014,Skelton2019}), or globally, \eg in long-term consequences of water-rock interactions such as "seismic degassing" (\eg \shortciteNP{Tamburello2018,Buttitta2020,Chiodini2020}).	
	
	In this study, we explore the hypothesis that seismic activity associated with strong earthquakes (magnitude $\geq$6) could actually result from the interplay between the solid Earth and the hydrosphere. To do so, we start by investigating the existence of a link between variations in global seismicity, sea level and planetary ephemerides. For this purpose, we analysed and compared the main pseudo-cycles detected and extracted from the global mean seismicity curve (M$\geq$6) since 1900, we later refer to it as the number of strong earthquakes (NSE), the length of day (LOD), and sea level at Brest tide gauge (SL$@$B), which is representative of the trends observed by worldwide tide gauges (\shortciteNP{Courtillot2022}). In section \ref{sec02} we present and discuss the dataset used, in section \ref{sec03} we compare the common pseudo-cycles extracted from all these geophysical data with planetary ephemerides, and in section \ref{sec04} we discuss our results and the proposed an earthquake trigger mechanism.
	
\section{\label{sec02} Data and Methods} 
	\subsection{Earthquake data}
	The Lisbon earthquake of 1755, undoubtedly the first earthquake to be documented with the rigor and standards of a contemporary scientific article, shows a remarkable singularity in the robustness of its observations (\eg \shortciteNP{Pereira1919,Baptista1998,Poirier2005,Poirier2006}). Unfortunately, not all seismic events around the world have been documented with the same quality. Most studies dealing with global seismicity start at the beginning of the 20th century and focus not on historical but rather on instrumental records even when they critically examine the available seismic catalogues (\eg \shortciteNP{Healy1992,Engdahl2002,Kagan2003,Albini2014,Kossobokov2020}).
	
\begin{figure}[H]
    \centering
    \includegraphics[width=0.8\textwidth]{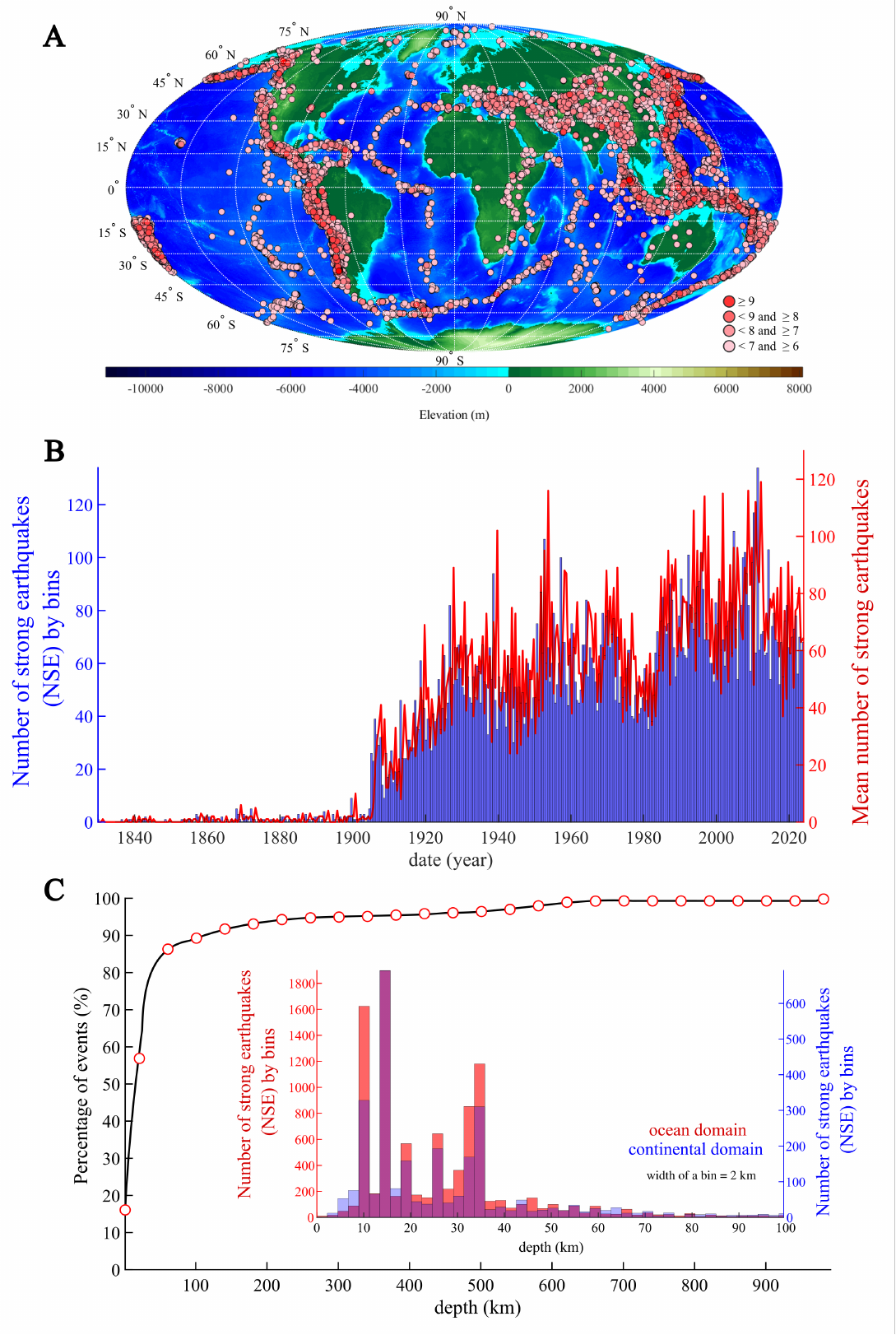}  
    \caption{(A) World map of seismic events with magnitudes of 6 or greater since 1900, from the USGS ANSS database. The shades of red indicate the earthquake magnitudes (B) Evolution of the number of strong earthquakes (NSE) over time since 1830, extracted from the USGS ANSS database. The 6 month-bin histogram is shown in blue and the 6-month moving average curve is shown in red. (C) Distribution of seismic events by depth. The black curve with red circles represents the cumulative percentage of the NSE as a function of depth. The event distribution for the first 100 km is shown in blue for the continents and in red for the oceans, both account for almost 90\% of the seismic events. The bins are 2 km width. }
    \label{fig:01}
\end{figure}

	In the present study, we have chosen to examine the temporal evolution since 1900, of all earthquakes with magnitude of 6 or greater from the U.S. Geological Survey (USGS) Advanced National Seismic System (ANSS) Comprehensive Catalog of Earthquake Events1. Figure \ref{fig:01}A shows the collection of these events. Not surprisingly, these earthquakes are found primarily, at plate boundaries where stresses and their variations are most pronounced according to higher levels of lithospheric blocks-and-faults hierarchy (\cf \shortciteNP{KeilisBorok1990}). The events in the centre of the map clearly delineate the contour of the African plate to the west, due to the spreading of the seafloor along the Mid-Atlantic Ridge, leading to the divergence of the African and South American plates. To the east, a similar phenomenon can be seen off the coast of the Indian Ocean, where the African and Australian plates diverge along the Indian Ocean Ridge. Similarly, but in a more continental domain, numerous and significant events occur to the north of the Arabian and Indian plates, both of which are in collision and convergence with the Eurasian plate. This interaction leads to the formation of the Zagros mountain range in Iran in the first case, and the Himalayan mountain range in the second case, as well as of numerous associated smaller continental blocks-and-faults in Eurasia. 
	
		The ANSS database contains about 14,000 seismic events  from January 01, 1900 to January 01, 2024 with magnitudes greater than or equal to 6. Figure \ref{fig:01}B shows a histogram of the number of strong earthquakes over time, with each bin representing a width of 0.5 years. We deliberately start at 1830 to illustrate the evident incompleteness of the database before the 20th century. It should be noted that before installation of the World-Wide Network of Seismograph Stations (WWNSS) in 1960s the list of even strong, magnitude 6 or larger earthquakes might be incomplete due to geographically inhomogeneous distribution of then available primarily not standarized seismographs. We computed the 6-month moving average of the number of strong earthquakes from 13,903 earthquakes  detected for the 1900-2024 period (Figure \ref{fig:01}B). This is this curve we will further analyze in this study and we refer to it as the number of strong earthquakes (NSE).
		
		It is also interesting, as a prelude to the mechanism we will propose at the end of our study, to present both the histogram and the cumulative depth distribution of all these earthquakes in Figure \ref{fig:01}C, which reproduces the classical observation of Gutenberg-Richter (1954). In black, we plot the cumulative percentage of the number of earthquakes as a function of their depth, which clearly shows that all earthquakes hypocenters fall within 10\%  of the Earth's radius from the surface, with almost 90\% of earthquakes with magnitude greater than or equal to 6 occurring within the first 100 km. If we look closely at the distributions of events down to 100 km (see the red and blue histograms in the same Figure \ref{fig:01}C), we see that most of them occurs above 40 km depth, with a median of about 18±10 km. Although this point is well known, it is important to recall it here. 
		
		To refine the analysis of the earthquake distribution, we have divided them into two categories depending on the epicenter location: the oceanic (red histogram) and continental (blue histogram) domains. It can be observed that the statistics are more or less the same, with the only notable difference being the factor of 2.5 between the number of seismic events under the oceans and on the continents, which ultimately corresponds to the different tectonic modes at plate boundaries.
	
	\subsection{Length of day}
	The length of day (LOD) refers to the duration of a terrestrial day, which is the time it takes the Earth to complete one full rotation on its axis. Although the nominal LOD is set at 24 hours, it can actually vary slightly due to various geophysical and astronomical factors. These variations in the LOD are mainly due to the gravitational interactions of the Moon and the Sun, which create tides in the oceans and the Earth's crust, slowing or speeding up the Earth's rotation. Mass movements within our planet, such as earthquakes, landslides, shifts in the Earth's core and changes in the distribution of mass in the oceans and atmosphere, also play a role in these variations. Atmospheric and oceanic effects, such as winds and ocean currents, can also transfer angular momentum to the Earth, changing its rate of rotation (\eg \shortciteNP{Melchior1958,Gross1997,Lambeck2005,Ray2014,LeMouel2019a}). 
	
\begin{figure}[H]
    \centering
    \includegraphics[width=\textwidth]{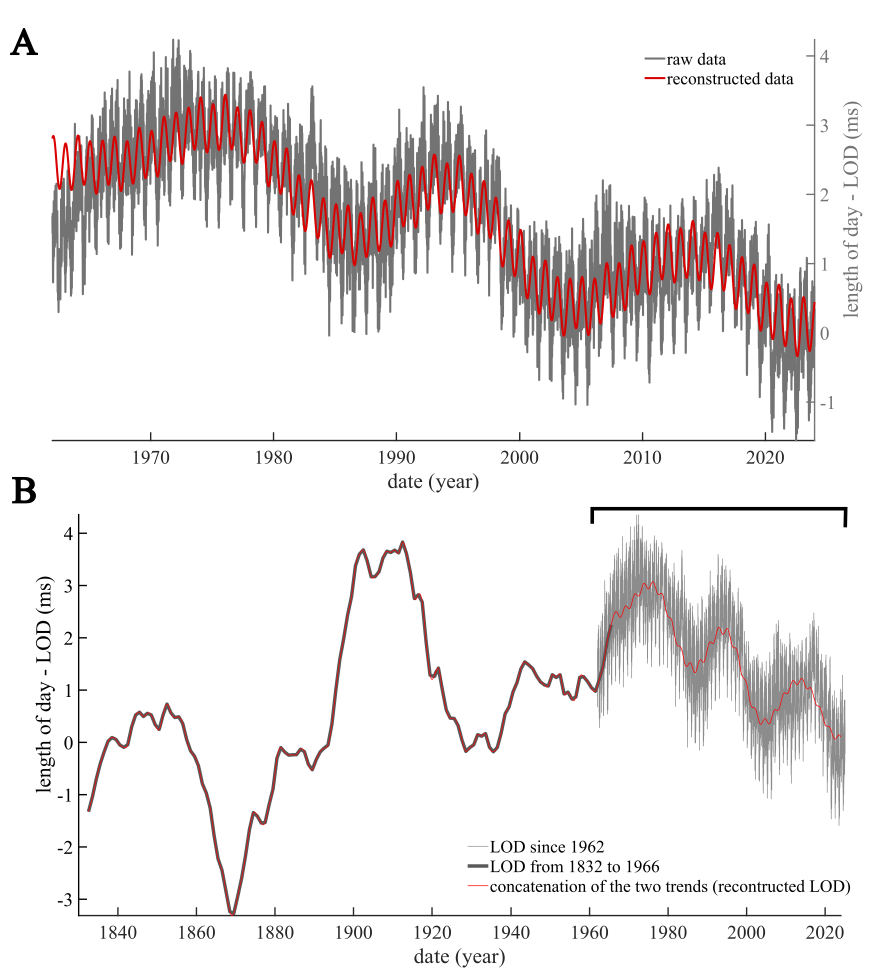}  
    \caption{ (A) The variation in the LOD (grey curve) provided by the EOP14C04 dataset since 1962. In red, the reconstructed signal based on the main long periods we extracted, e.g trend, $\sim$19 years, QBO, $\sim$1 year, which represents 80\% of the originail signal. (B) Monthly values of the LOD data (LUNAR97, 1832-1997; represented by the bold black curve) obtained from Stephenson and Morrison (1984) and Gross (2001), alongside daily values (1962-present, represented by the grey curve) provided by IERS. The red curve represents the reconstruction of the LOD time-series since 1832 using the data from Stephenson and Morrison (1984) and Gross (2001) (bold black curve) and the LOD measurements from 1962  made by IERS.}
    \label{fig:02}
\end{figure}

	Although variations in the LOD are generally very small, on the order of milliseconds, they can be measured with great precision using a combination of modern geophysical and astronomical techniques sensitive to the Earth's gravitational field, based on observations of stars and satellites such Very Long Baseline Interferometry (VLBI), Satellite Laser Ranging (SLR), and gravimetric measurements (for more details, see \shortciteNP{Bizouard2019}).	Measurements of the variations in the  LOD are maintained by the International Earth Rotation and Reference Systems Service (IERS), which produces the EOP14C04 dataset that we have analysed (Figure \ref{fig:02}A). This dataset covers the period from 1 January 1962 to 11 March 2024 with a daily sampling interval. 
	
	It is, however, possible to extend our analysis further back in time, as other approaches have been used to provide more constraints on the variations in the LOD over past centuries. Thus, for longer durations that are more compatible with the seismic event series (Figure \ref{fig:01}), we rely, as in the past (\cf. \shortciteNP{Lopes2022a}), on the compilations of  lunar occultation, optical astrometric, and space-geodetic measurements of the Earth’s rotation whose analyses by \shortciteNP{Stephenson1984} and \shortciteNP{Gross2001} resulted in a long LOD time-series which covers the period from 700 BC to 1980 AD. Figure \ref{fig:02}B shows the IERS EOP14C04 series from 1962 to the present (grey curve), and the \shortciteNP{Stephenson1984} series from 1832 (black curve). In red is the concatenation of the low frequency periods, \ie trend, QBO, 19-years and 1-year periods, extracted from EOP14C04 by Singular Spectrum Analysis (SSA, see section "The pseudo-cycle extraction method") with periods compatible with the frequency support of long LOD time-series from  \shortciteNP{Stephenson1984} and \shortciteNP{Gross2001}. Over the $\sim$20 year of overlap of the long and short LOD time-series, we calculated their average. The red curve (Figure \ref{fig:02}B) thus represents the LOD reconstruction since 1832.

	\subsection{The Brest tide gauge}
	The Brest tide gauge is unique in that it is both the oldest tide gauge in the world, having been commissioned in 1807 and being still in operation (\cf \shortciteNP{Woppelmann2006,Woppelmann2008}), and, at the same time, the sea-level trend it records appear to be  representative of those  observed by all tide gauges in the northern hemisphere (\shortciteNP{Nakada2005,Courtillot2022}). Moreover, all these tide gauges show a very similar content of periodic components that is also observed globally, and that is consistent with variations of global pressure data (\shortciteNP{Courtillot2022}). More, these oscillations are also found in the movement of the mean Earth's rotation pole (\cf \shortciteNP{LeMouel2021}). These observations are most likely due to the particularly stable bathymetry over tens of kilometres of the Breton seafloor (France).
	
\begin{figure}[H]
    \centering
    \includegraphics[width=\textwidth]{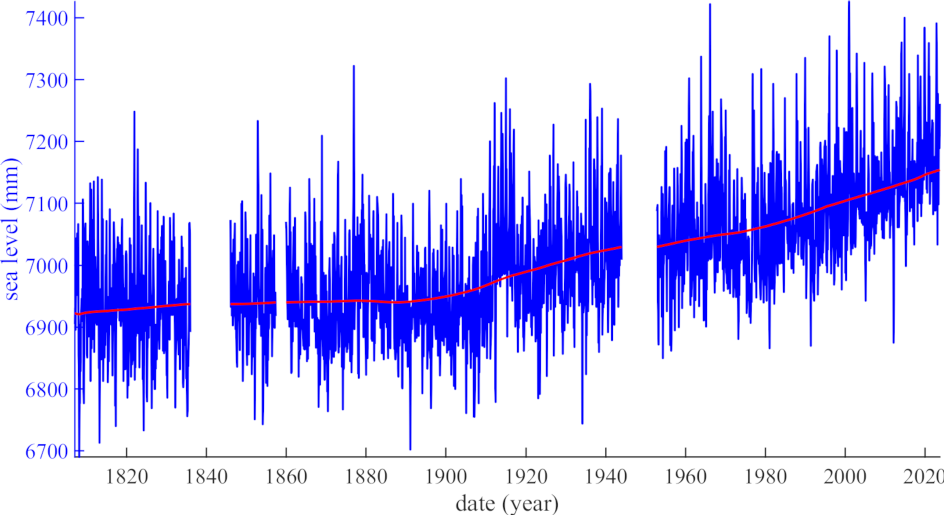}  
    \caption{The sea level recorded at Brest since 1807. The data gaps correspond to wartime periods when the tide gauge was destroyed. In blue are the raw data, in red, the mean trend of sea level.}
    \label{fig:03}
\end{figure}

	In Figure \ref{fig:03} we have plotted the sea level recorded by this tide gauge, using monthly data (blue curve) provided by the Permanent Service for Mean Sea Level (PSMSL, https://psmsl.org/, \shortciteNP{PSMSL2024,Holgate2013}). As can be seen, measurements at Brest were interrupted for some time in the 1840s and 1940s, when the tide gauge was destroyed during the wars.
	
	\subsection{The Sun, the Moon and the planet’s ephemerides}
	The ephemerides of the Moon, the Sun and planets are calculated by the Institut de Mécanique Céleste et de Calcul des Ephémérides (IMCCE, https://www.imcce.fr/). We determined, in a geocentric frame, the variations of the positions according to the declination of all the celestial bodies (\eg Moon, Sun, Jupiter, Saturn \etc).  
	
	\subsection{The pseudo-cycle extraction method}
	The data we analysed are not necessarily stationary in the strict sense, which may be expected considering the nature of the phenomena, and can sometimes be discontinuous, as is the case for the sea level in Brest. We therefore need a sufficiently robust method to extract pseudo-cycles. By pseudo-cycle we mean an oscillation that can be highly modulated in both phase and amplitude, but whose Fourier spectrum, although spread around a nominal frequency, remains exclusively centred around that frequency. Singular Spectrum Analysis (SSA) is an ad hoc decomposition method, i.e. the orthogonal basis on which the signal information is projected and is constructed from the information of the signal itself (unlike the infinite sines of the Fourier transform for example), which makes it particularly well suited to this problem (\cf \shortciteNP{Vautard1989,Vautard1992}).
	
	We have already described the SSA algorithm in \shortciteNP{Lopes2022b}, and it is even the subject of a reference work by \shortciteNP{Golyandina2013}, which details all its possible variations. In the present paper we briefly outline the main aspects of the method: it consists of four steps. In the first, embedding step, the data are projected into a specific matrix, either a Toeplitz matrix or a Hankel matrix (\cf \shortciteNP{Lemmerling2001}). Essentially, a segment of the signal of length $L$ is written in each column of the matrix. The value of $L$ determines the physical properties (\eg chaotic, strange attractor, short or long period, \etc) that will be extracted from the data. The second step consists in diagonalising the previously matrix built, typically by Singular Value Decomposition (SVD, \shortciteNP{Golub1971}), which yields two matrices of eigenvectors inducing a passage from the data to the dual space (thus an ad hoc orthogonal basis), with the eigenvectors sorted in descending order of the corresponding eigenvalues. This is analogous to a Fourier spectrum. In the third step, known as the grouping step, similar or close eigenvectors and eigenvalues are paired, known as the grouping step. Finally, in the last step, called Hankelization, the grouped eigenvector/eigenvalue pairs are returned to the data space.  That is how the pseudo-cycles that correspond to these pairs are extracted and reconstructed.	
	
	For each component extracted in the different geophysical time-series, we calculated the associated variance. This latter corresponds to the square root of the sum of the squares of the eigenvalues that comprise the component. We refer to the total variance or total energy (of the originial signal) when calculating the variance for all eigenvalues.
	
	One must keep in mind that the earthquake catalog is not uniformely complete, especially for magnitudes above 6, for the time period from 1900 to 2024 (\eg \shortciteNP{Engdahl2002,Kossobokov2006,Kossobokov2018}). However, we are studying the number of strong earthquake as a scalar at the planetary scale. Thus, the associated error remains consistent over time and across locations, which tends to be reduced over recent time due to instrumentation increase. Therefore, the shapes of variations and pseudo-periodicities detected and extracted should be similarly affected, allowing us to perform a reliable analysis.
	
	\section{\label{sec03} Results and descriptions of the analysed signals}
	The main long-period pseudo-cycles ($\geq$ 1 year) that we have detected and extracted are  shown in Supplementary Material (Sections  \ref{supp:A} to \ref{supp:C}); only the trends are not shown for the sake of  parsimony. The analysis highlights the coincidence of 7 cycles between the NSE and the LOD, and 4 cycles with the variation of the SL@B (Table \ref{table:01}). These pseudo-cycles are, in ascending order: the 1-year seasonal oscillation  the 2.3-year Quasi-Biennial Oscillation (QBO, \shortciteNP{Baldwin2001}), a $\sim$11-year pseudo-cycle that may have a connection with Jupiter’s orbit, except for the SL@B, where it was absent – unlike in the mean global sea level,  \eg \shortciteNP{Courtillot2022,LeMouel2021,Lopes2021}) –  a $\sim$14-year pseudo-cycle (periodicity related to Jupiter + Uranus, \shortciteNP{Scafetta2022}), a $\sim$19-year pseudo-cycle corresponding to the precession of the lunar orbit which has an (exact) period of 18.6 years, a $\sim$33-year pseudo-cycle, and finally a $\sim$65 year pseudo-cycle. In Supplemenaty material Figures \ref{supp:A}5, \ref{supp:B}5 and \ref{supp:C}4, we illustrated some of these pseudo-oscillations and their reconstruction for the three time-series analyzed. The reconstruction of the extracted pseudo-periods shows that the sum of these cycles corresponds to about 88\% of the NSE, about 89\% of the variation in the LOD and about 90\% of the variation in the SL@B. These pseudo-periods are summarised in Table \ref{table:01}. One can note the high levels of uncertainty in the determination of long-term cycles: for the$\sim$33-year cycle, a factor about 1.3 between the NSE and the SL@B,  and similarly, a factor of 0.4 for the $\sim$60-year pseudo-cycle. In the following, we present the common cycles extracted in the three geophysical time-series.	
	
\begin{table}[htbp]
  \centering
  \begin{tabular}{p{4cm}|p{4cm}|p{4cm}}
    \hline\hline
     Number of strong earthquakes (NSE, years) & Length of day (LOD, years) &  Brest Sea-Level (SL@B, years) \\
	\hline\hline
    $1.00 \pm 0.00$ & $1.00 \pm 0.01$ & $1.00 \pm 0.01$ \\
    $2.35 \pm 0.03$ & $2.36 \pm 0.06$ & \\[1ex]
    $11.36 \pm 0.57$ & $11.80 \pm 1.21$ & \\[1ex]
    $14.48 \pm 1.03$ & $13.48 \pm 0.71$ & $13.40 \pm 0.70$ \\
    $18.66 \pm 2.04$ & $19.23 \pm 3.79$ & $18.94 \pm 3.34$ \\[1ex]
    $34.10 \pm 43.33$ & $33.08 \pm 3.85$ & $36.55 \pm 13.05$ \\
    $65.03 \pm 22.25$ & $65.55 \pm 17.02$ & \\
    \hline\hline
    $\sim 88\%$ & $\sim 89\%$ & $\sim 90\%$ \\
    \hline\hline
  \end{tabular}
  \caption{Summary of pseudo-periods detected and extracted from three geophysical datasets, \ie NSE, LOD, and SL@B. The last line indicates the percent of signal reconstructed using the listed cycles.}
  \label{table:01}
\end{table}

\begin{figure}[H]
    \centering
    \includegraphics[width=\textwidth]{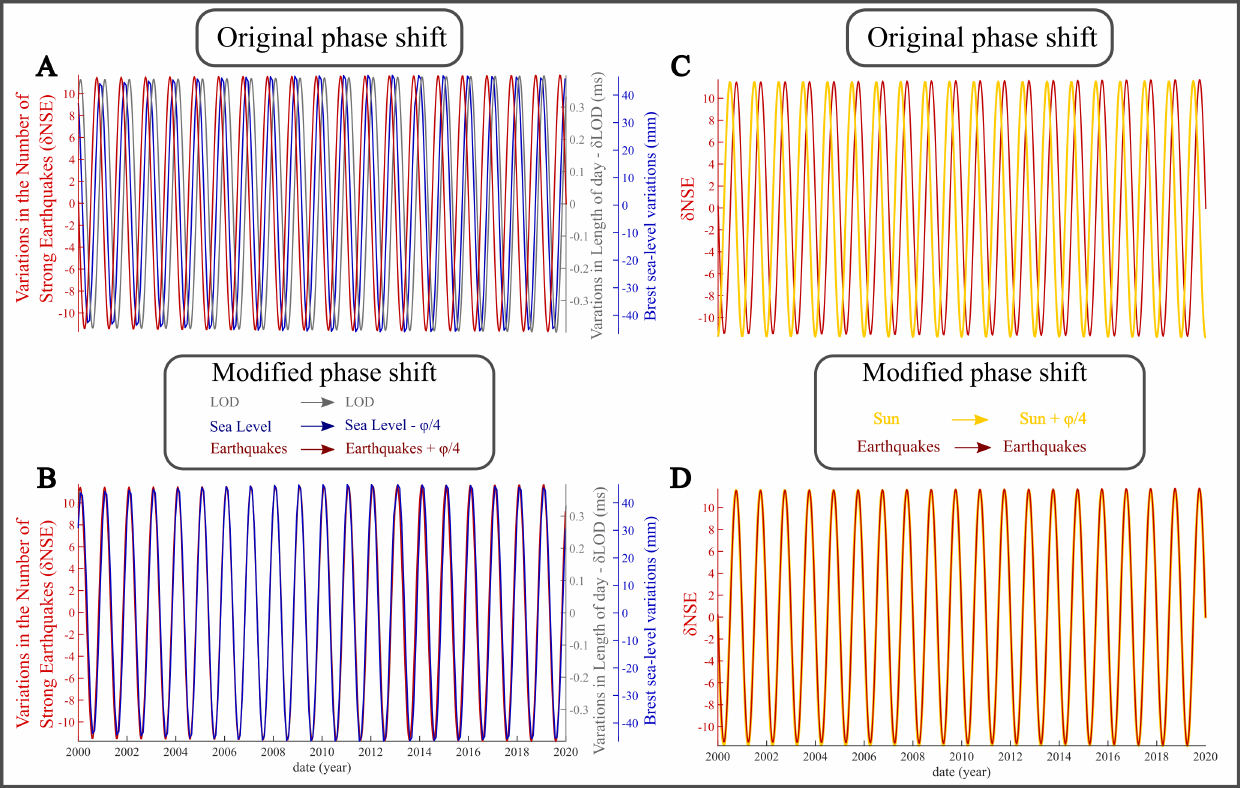}  
    \caption{The seasonal oscillation. (A-B) The $\sim$1-year pseudo-cycle extracted by SSA since 2000 for better readability, from the NSE (red curve), LOD (grey curve) and SL@B (blue curve). (C-D) The $\sim$1-year pseudo-cycle extracted from the NSE (red curves) superimposed on the same component but extracted in the ephemeride of the Sun (yellow curve). In (A) and (C), the original phase shifts extracted by SSA are shown; in (B) and (D), the same curves as in (A) and (C) respectively, are phase shifted by one quadrature (± $\varphi$/4) for the NSE and SL@B in (B), and the Sun (D). }
    \label{fig:04}
\end{figure}

	\paragraph{seasonal oscillation}
	Figure \ref{fig:04}A shows the seasonal pseudo-cycle extracted from the three geophysical datasets. The time axis starts from 2000 for better readability of the observations. As we can see, the variations in the SL@B (blue curve) and the NSE (red curve) are in perfect phase quadrature (± $\varphi$/4), which corresponds to a temporal derivative, with the variations in the LOD. Once this phase shift is applied, all curves overlap almost perfectly (Figure \ref{fig:04}B).

 In Figure \ref{fig:04}C we have superimposed the ephemeride of the Sun (yellow curves) on the annual variation in the NSE over the same period. A phase quadrature is clearly visible, similarly as between the SL@B and NSE and LOD (Figure 4A-B) which, when applied to the envelope, perfectly matches the geophysical observations (\cf Figure \ref{fig:04}D).
 
\begin{figure}[H]
    \centering
    \includegraphics[width=\textwidth]{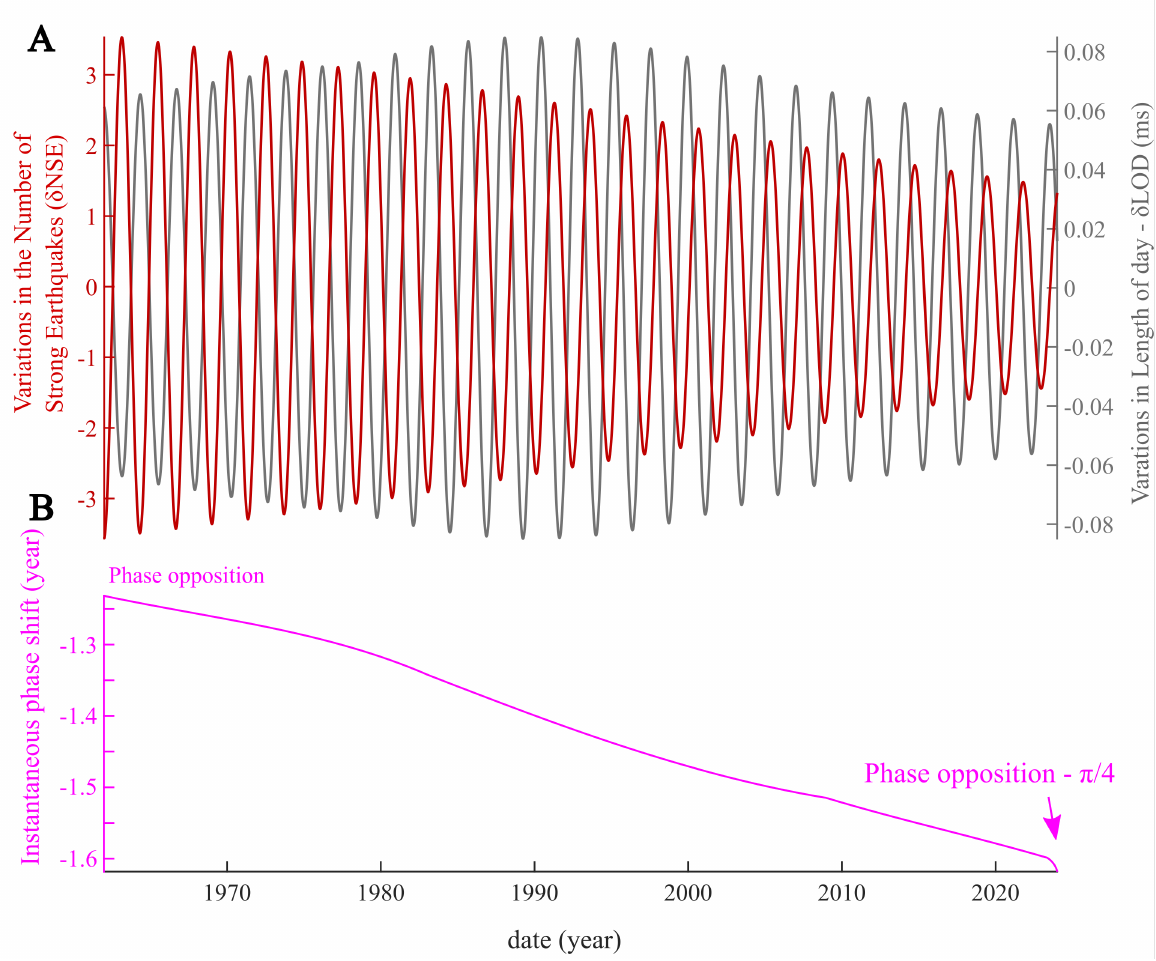}  
    \caption{The Quasi-Biennial Oscillation (QBO) of $\sim$2.3 years. (A) The QBO extracted from the NSE (red curve) is superimposed on the QBO extracted from the LOD (grey curve). (B) The evolution of the instantaneous phase shift since 1962; over 60 years, a phase opposition transitions to a phase quadrature ($\sim$ -$\pi$/4).}
    \label{fig:05}
\end{figure}

	\paragraph{The Quasi-Biennial oscillation}
	As mentioned above, we have not identified the QBO in the SL@B, but this does not mean that it does not exist in general. In Figure \ref{fig:05}, we show the excellent agreement between the periods of the QBO extracted in the NSE (red curve) and the QBO from the LOD measurements (grey curve). Clearly, over the period of interest here, \ie 1962 to 2024, which coincids also with the period of the improved earthquake hypocenter determinations after installation of WWNSS, the phase variation between the two geophysical parameters does not appear to be constant, as was the case for the forced seasonal oscillation (see Figure \ref{fig:04}). For this reason, we have evaluated and plotted the evolution of the instantaneous phase shift over time in Figure \ref{fig:05}B. As we can see, we start in 1962 with an almost perfect phase opposition between the two geophysical measurements and arrive in 2024 to a simple quadrature.   
	
\paragraph{The $\sim$11-year pseudo-cycle}	
	 It has long been known that one of the most significant pseudo-cycle components in the LOD occurs at a period of about 11 years (\cf \shortciteNP{Stephenson1995,LeMouel2019a}), and might well express the influence of Jupiter. This pseudo-cycle has also been identified in the polar motion (\cf \shortciteNP{Lopes2017}), although its amplitude is more modest. Surprisingly, it is found only occasionally and modestly (\eg \shortciteNP{Currie1981}) or not at all (\cf \shortciteNP{Courtillot2022}) in the global mean sea level, which may seem paradoxical when considering the mechanisms that should accelerate or decelerate the Earth's rotation according to Laplace’s theory (\shortciteNP{Lopes2021}).		
	 
\begin{figure}[H]
    \centering
    \includegraphics[width=\textwidth]{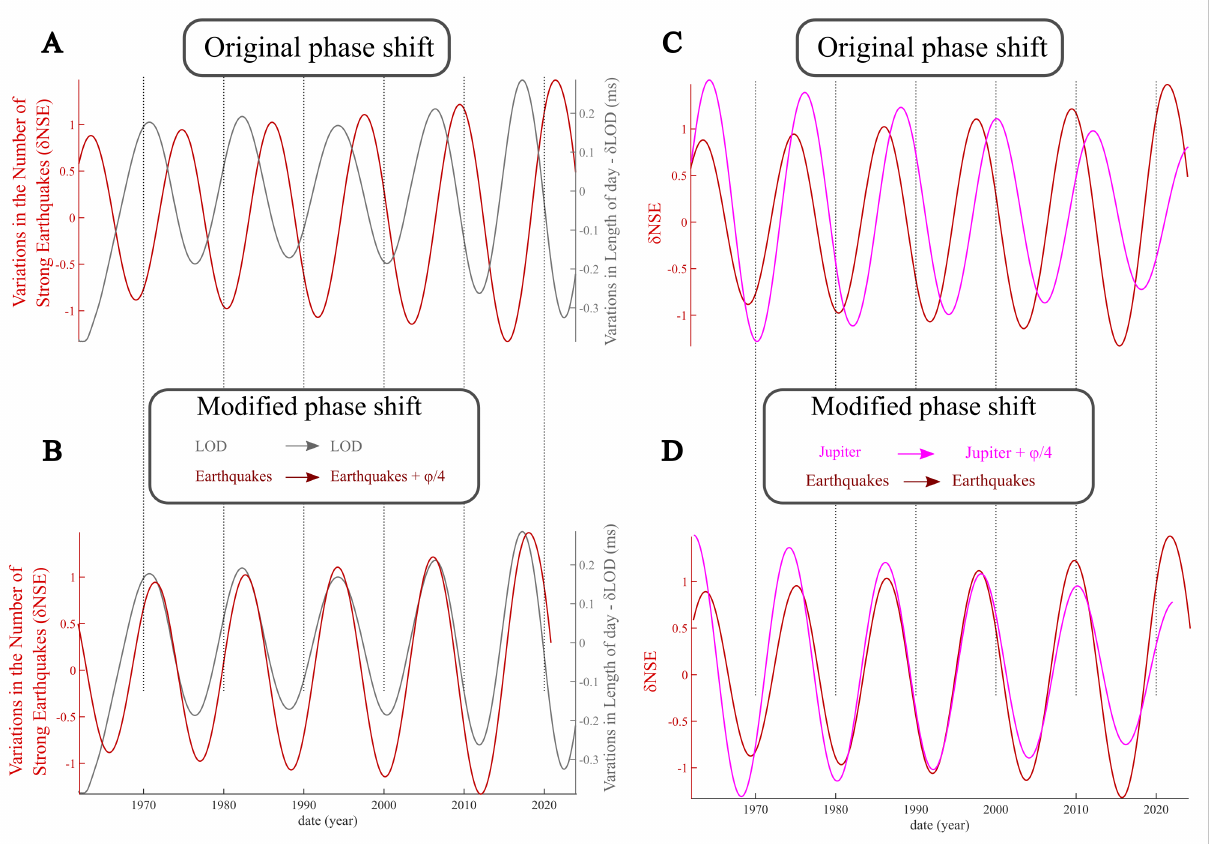}  
    \caption{ The $\sim$11-year pseudo-cycle. (A-B) The $\sim$11-year oscillation detected and extracted from the  NSE (red curve) and LOD (grey curve) since the 1960s. (C-D) The same pseudo-cycle (red curves) superimposed on the same component but extracted in the ephemeride of Jupiter (purple curve). In (A) and (C) are shown the original phase shifts, in (B) and (D) are the same curves as in (A) and (C) respectively, but phase shifted by one quadrature (±$\varphi$/4).  }
    \label{fig:06}
\end{figure}

	 We have extracted the 11-year pseudo-cycle from the NSE and plotted  it in Figure \ref{fig:06}A (red curve). In Figure \ref{fig:06}B, we have shifted the seismic oscillation by one phase quadrature, as we did previously in Figure 4B for the seasonal pseudo-cycle. Once again, the two curves, LOD and NSE, are in phase. In Figure \ref{fig:06}C-D we compare the Jupiter ephemerides (pink curve) with the seismic pseudo-cycle, and again the phase match is perfect once the quadrature shift is removed.
	 
\begin{figure}[H]
    \centering
    \includegraphics[width=\textwidth]{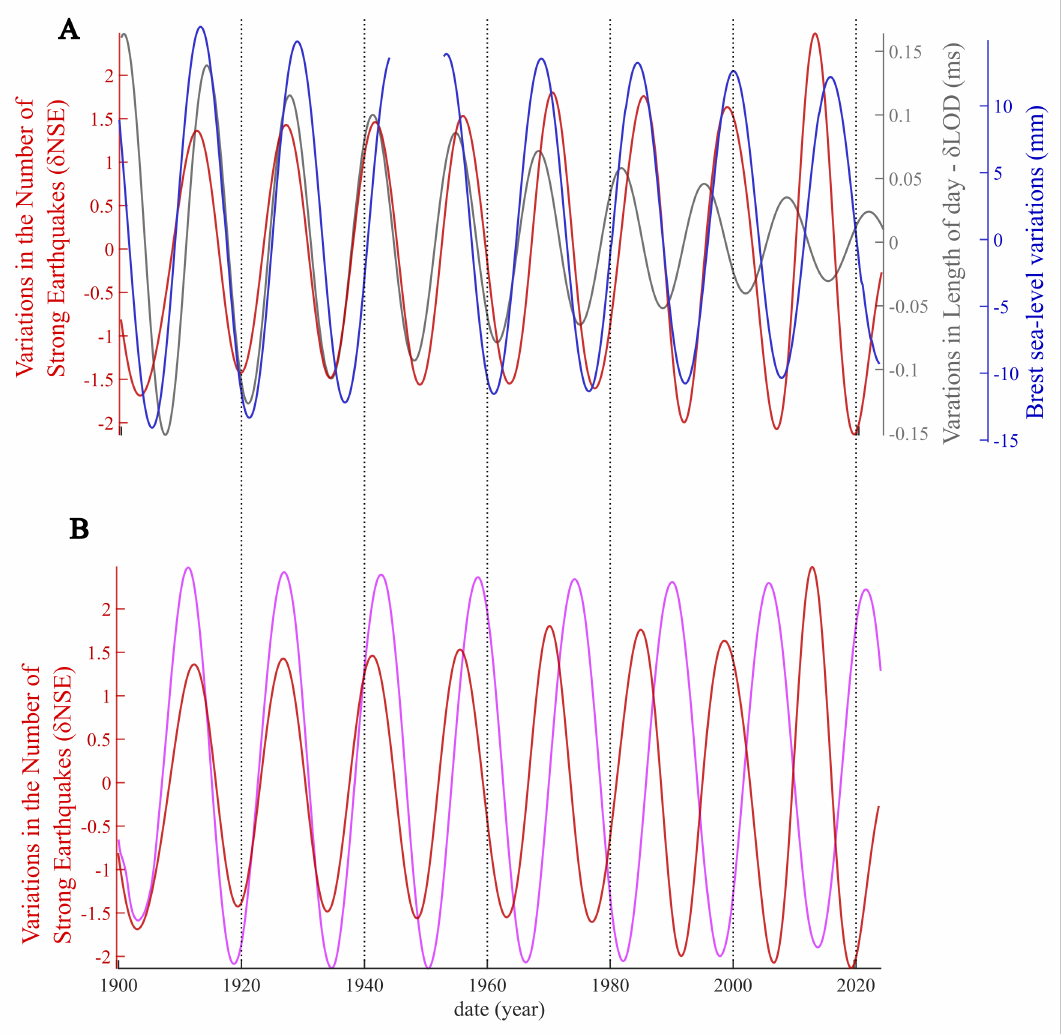}  
    \caption{The $\sim$14-year pseudo-cycle. (A) The 14-years oscillation extracted from the three geophysical data sets without phase shifting with in red that of the NSE, in blue the SL@B and the reconstructed LOD in grey; (B), we reported the same component of the NSE (red curve), together with that extracted from the ephemerides of Jupiter and Uranus (purple curve).}
    \label{fig:07}
\end{figure}

	 \paragraph{The $\sim$14-year pseudo-cycle}
	 Unlike previous cycles, this one shows significant phase modulation on a century scale (Figure \ref{fig:07}A). Seismic activity (NSE) and SL@B, which are almost in phase during the 120 years of observation, were in phase with the reconstructed LOD at the beginning of the last century, and reached a phase opposition only by the mid-2010s (Figure \ref{fig:07}A). This consistent phase shift may have several causes, but it is most likely that the primary cause taking place at this time scale is that the extracted frequencies are slightly different (see the associated uncertainties in Table \ref{table:01}), although their associated spectral widths make them compatible. An interesting observation is that the amplitude modulations between seismic activity and sea level also appear to exhibit an inverse correlation over the entire period. 	
	 
	 The planetary commensurability corresponding to this period is generally attributed to the Jupiter + Uranus pair (\cf \shortciteNP{Scafetta2022}). We have superimposed the evolution of this commensurability (Laplace’s resonnance) using the ephemerides of the two aforementioned planets (purple curve) on the curve of the NSE (red curve) in Figure 7B. A good agreement can be observed during the period 1900-1950, after which the gradual phase shift between the two physical phenomena becomes more pronounced, and since the early 2000s the two curves are in phase opposition.This transition might result from the aforementioned revolutionary change in earthquake determinations in the 60’s (see subsection Earthquake data). 	
	
\begin{figure}[H]
    \centering
    \includegraphics[width=\textwidth]{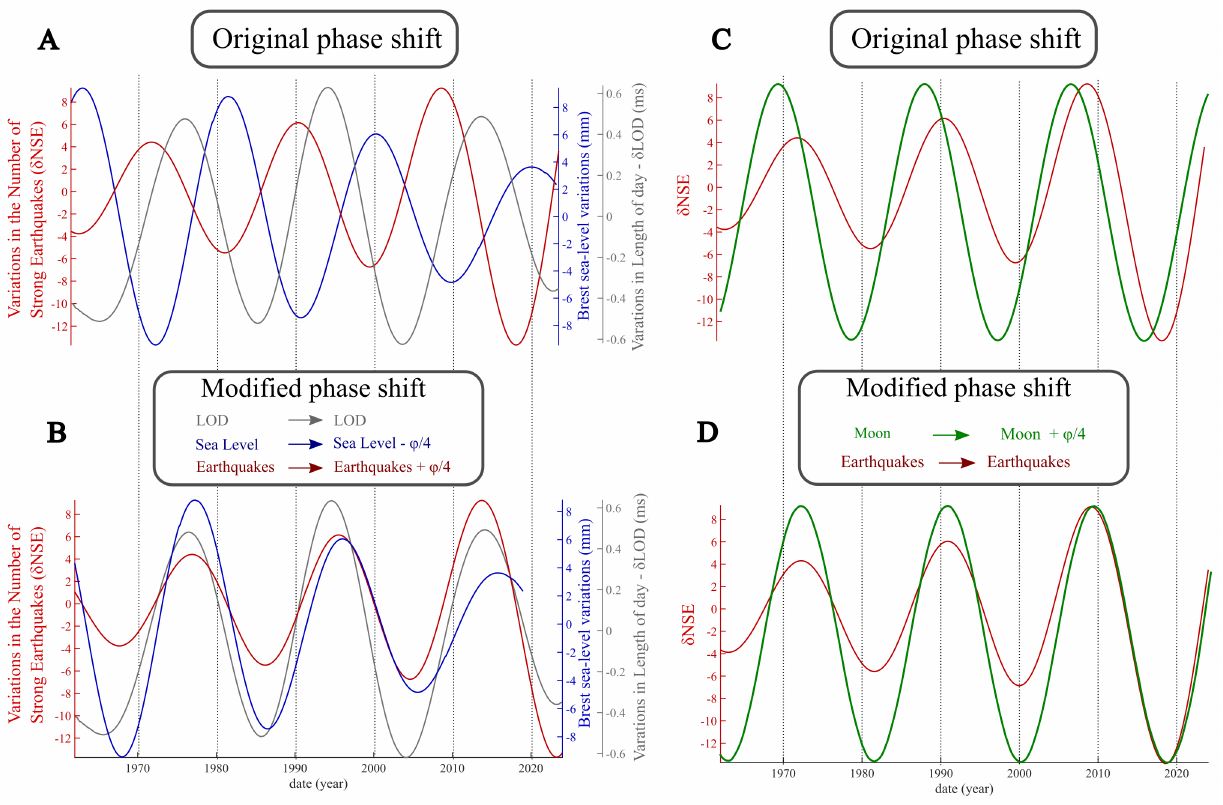}  
    \caption{The 18.6 year pseudo-cycle. (A-B) The $\sim$19-year component extracted in the NSE since 1962 (red curve), LOD (grey curve) and the SL@B (blue curve). (C-D) The $\sim$19-year component extracted in the NSE (red curve) represented together with that extracted from the Earth-Moon distance (green curve).  In (A) and (C) are shown the original phase shifts, in (B) and (D) are the same curves as in (A) and (C) respectively, but phase shifted by one quadrature (± $\varphi$/4).  }
    \label{fig:08}
\end{figure}

\paragraph{The 18.6-year lunar pseudo-cycle}
	The $\sim$19-year pseudo-period detected and extracted from the three geophysical records corresponds to the 18.6 year oscillation, which, as mentioned above, is caused by the precession of the lunar orbital plane. Figure 8A shows the corresponding curves. As in the case of the annual oscillation (\cf. Figure \ref{fig:08}A), the SL@B and NSE are both in phase quadrature with the variations in the LOD. Once this phase quadrature is applied, all geophysical records are almost perfectly in phase (\cf Figure \ref{fig:08}B). The ephemeride of the Moon (green curve) and the NSE (in red) in Figure 8C are almost perfectly in phase, when we shift it by a phase quadrature (Figure \ref{fig:08}D). 
	
\begin{figure}[H]
    \centering
    \includegraphics[width=\textwidth]{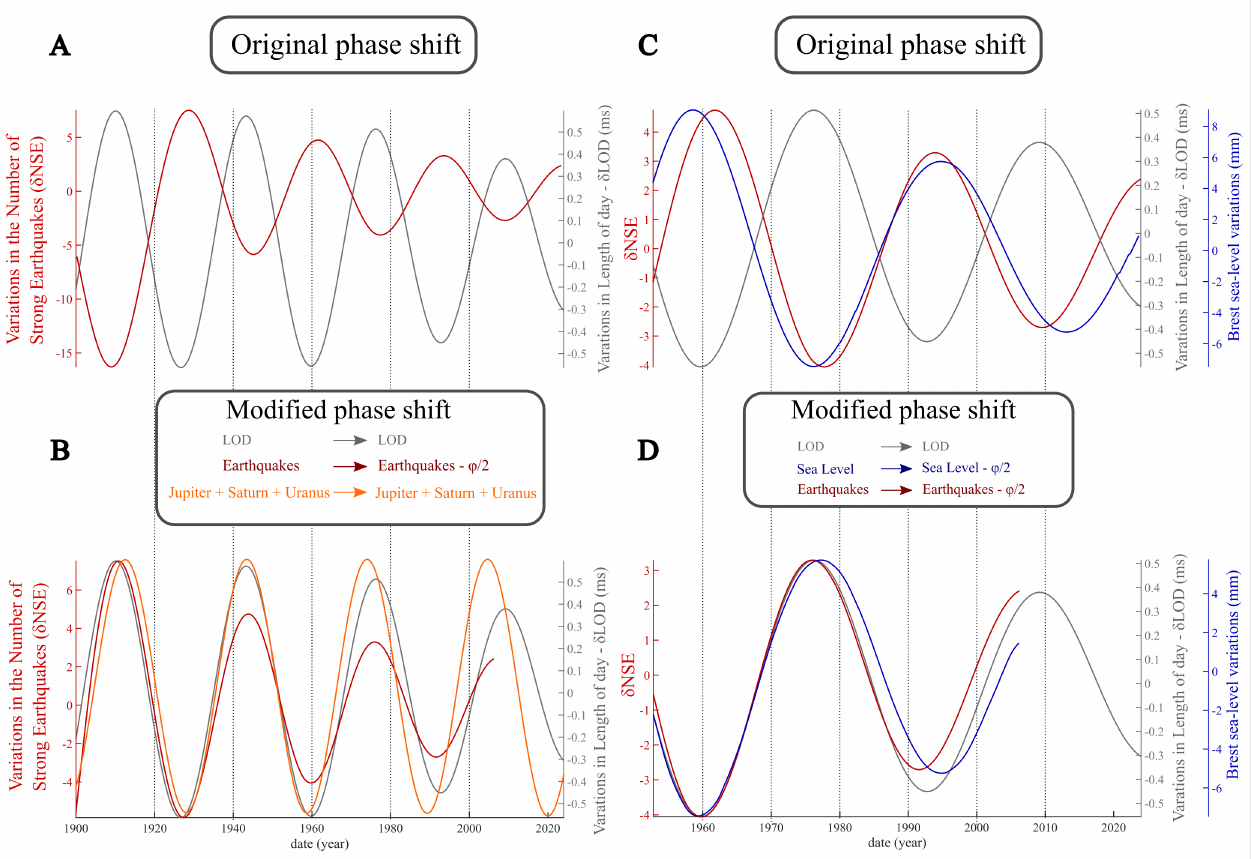}  
    \caption{The $\sim$33-year pseudo-oscillation. (A-B) The 33-year component extracted from the NSE since 1900 (red) and from reconsructed LOD (grey). In (B), the ephemerides of the combination Jupiter + Saturn + Uranus was added. (C-D) Same component but shown from 1958, for three geophysical time-series: the NSE (red), the SL@B (blue) and the reconstructed LOD (grey).  In (A) and (C) are shown the original phase shifts, in (B) and (D) are represented the same curves, but phase shifted by one phase opposition (±$\varphi$/2).  }
    \label{fig:09}
\end{figure}

\paragraph{The $\sim$33-years pseudo-cycle}
	Although SSA is more robust for extracting components than, for example, wavelet methods, it is of course not perfect. As shown in Supplementary Material Figure \ref{supp:C}1 for the Brest tide gauge analysis, data gaps lead to edge effects on the waveforms that are not insurmountable, but still present. To avoid misinterpretation due to these edge effects, Figure \ref{fig:09}A shows the superposition of the 33-year pseudo-cycles of the reconstructed LOD (grey curve) and seismic events (red curve) since 1900. A phase opposition appears which, when applied to seismicity, results in an almost perfect superposition of the two geophysical records (Figure \ref{fig:09}B). Figures \ref{fig:09}C-D show the same curves but starting in 1958, which is the beginning of the last continuous segment of the Brest tide gauge record. Thus, in Figure \ref{fig:09}C, we have added the 33-year oscillation of SL@B (blue curve) to the two previous geophysical records. Again there is a phase opposition with the reconstructed LOD which, when removed from the SL@B, allows a good superimposition of the three geophysical records (Figure \ref{fig:09}D).
	
	One can note that this 33-year cycle is also found in sunspots (\eg \shortciteNP{Usoskin2017,LeMouel2020}) and corresponds to one of Laplace's commensurable ratios between Jupiter, Saturn and Uranus (\eg \shortciteNP{Morth1979}). Despite these observations, it is important to keep in mind that this $\sim$33-year pseudo cycle, is one of those that are less reliable, with the $\sim$60 years. It is the unique long oscillation with uncertainties larger than the periodicity itself (Table \ref{table:01}).
	
\begin{figure}[H]
    \centering
    \includegraphics[width=\textwidth]{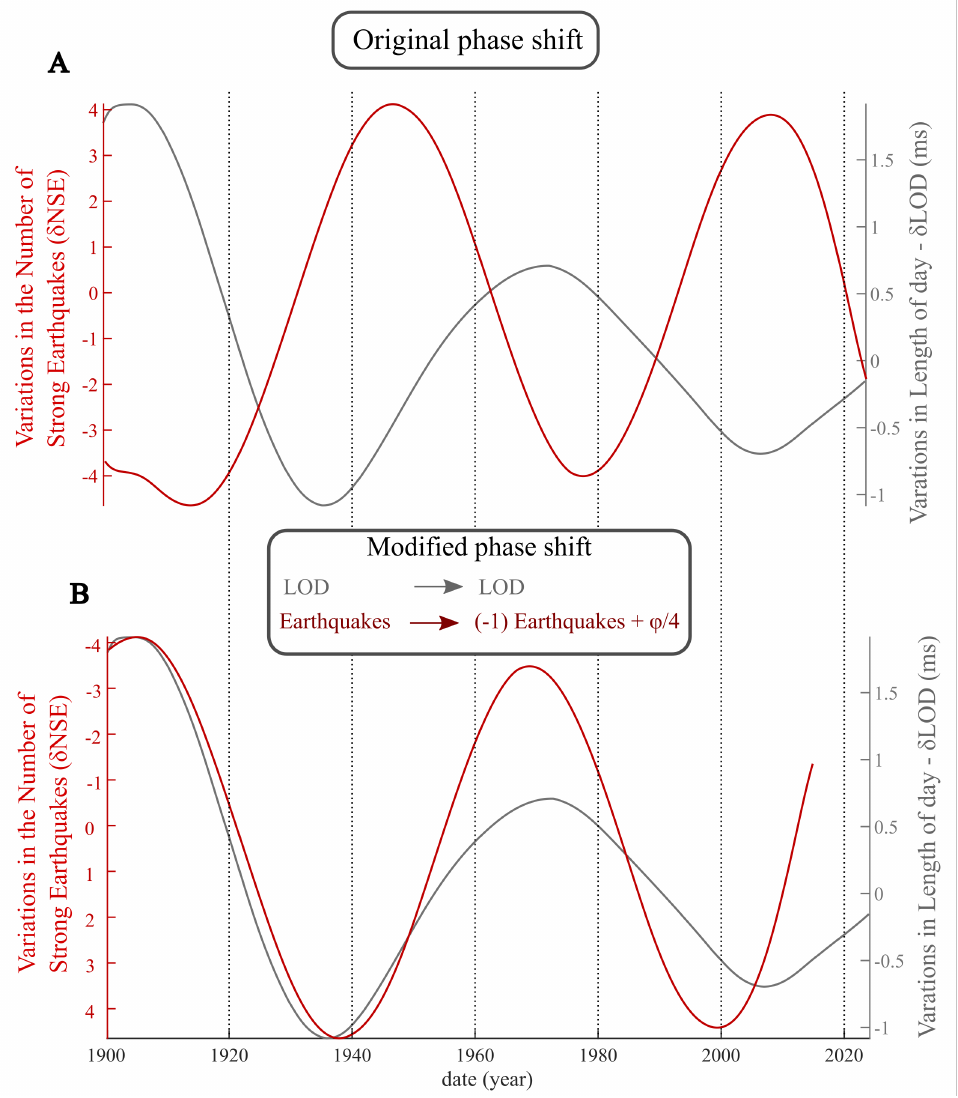}  
    \caption{The $\sim$60-year pseudo-cycle, with the reconstructed LOD (grey curve) and NSE (red curve) shown (A-B). In (B), the same curves are realigned by a phase quadrature.}
    \label{fig:10}
\end{figure}

\paragraph{The $\sim$60 year pseudo-cycle}	
	Regarding this last pseudo-cycle of $\sim$60 years, whose determination is less reliable if we consider the associated uncertainties (Table \ref{table:01}), we did not detect it in the Brest sea level data (SL@B),  which does not necessarily mean that it does not exist, since we know that it clearly appears in the global mean sea level, with a period of 57.5 ± 7 years (\shortciteNP{Courtillot2022}, Table 2). As for its correspondence with ephemerides or commensurabilities, one can note that this cycle corresponds to the Uranus + Neptune pair, although this observation should be taken with caution. Only barely two cycles have been observed since 1900 (Figure \ref{fig:10}), and like the trend, which we have not addressed here, one cannot reasonably say more than that the $\sim$60 year pseudo-cycle is found in the mean sea-level, the mean seismicity and the reconstructed LOD .
	
	In summary,  six of the seven pseudo-cycles detected and extracted, are in perfect quadrature or opposition and they appear in the ephemerides of the Moon, Sun and the main planets of our solar system, which are known to influence the rotation of the pole (\shortciteNP{Lopes2021}). Moreover, it is the first time that the QBO, which is not in quadrature nor in opposition, has been unambiguously detected and extracted from global seismicity.
	
\section{\label{sec04}Discussion - Conclusions}	
	In this study, we have investigated the existence of periodic components, in the NSE over an interval of 124 years, starting in 1900. We detected seven cycles, spanning interannual to decadal periods, common to those present in the polar motion, and more specifically in the LOD; four of them, being also present in the sea-level (Table \ref{table:01}). These observations align with studies claiming that strong earthquake occurrence appears to follow lunisolar tidal cycles for short periods (\shortciteNP{Lockner1999,Metivier2009,Petrosino2018,Varga2019,Zuddas2024}), and solar activity for longer ones (\shortciteNP{Simpson1967,Mazzarella1989,Choi2010,Marchitelli2020}). Despite these observations and the fact that a sample of earthquakes seems to follow specific periods, neither plausible physics-based mechanism has been proposed, nor an accurate relationship has been established allowing to predict earthquake occurrence (\eg \shortciteNP{Tsuruoka1995,Kossobokov2020}). The seven cycles we identified in the NSE are known as commensurable periods corresponding to specific orbital configuration, \eg alignment or quadrature of individual or paired planets of the Solar system  (\shortciteNP{Morth1979,Courtillot2021,Lopes2021,Scafetta2022}). We interpret their presence in the occurrence of strong earthquakes at global scale as an astronomical forcing which likely acts as a trigger through the transfer of energy via angular momentum exchanges as proposed by previous studies  (\shortciteNP{Courtillot2021,Lopes2021,LeMouel2023}) and which influences in one way or another rock failure. However, it is well known that the strongest tides on Earth, that are of lunisolar origin, are associated with tidal stresses that are 2-4 orders of magnitude lower than the strength of rock and faults ($\sim$101–103 MPa, \shortciteNP{Scholz2019}). The influence of other planets of our solar system is therefore weaker than lunisolar ones, raising the question on how such interaction can occur.
	
	For decades, seismic activity has been associated with degassing but also changes in the groundwaters including their composition, which has stimulated extensive researches for the early detection of geochemical precursors of earthquakes (\eg \shortciteNP{Teng1980,King1996,Toutain1999,Chiodini2004,Perez2008,Woith2015,Tamburello2018,DiLuccio2022,Skelton2014,Skelton2019}). Moreover, some studies have suggested that the nature of the faults and therefore the tectonic regime, could explain the degassing patterns observed around the world (\shortciteNP{King1993,Tsuruoka1995,Tamburello2018}). However, if we consider global seismicity as an ergodic phenomenon, the analyses should be generally consistent and independent of the geological nature of the area where an earthquake struck. This idea is supported by systematic observations we made in our analysis of extracted periodic components. Thus, the phase lag of $\varphi$/4  – $\varphi$ the cycle, was observed between (i) most of the components extracted in the three geophysical time-series studied, \eg the NSE, SL@B and LOD, on one hand, and (ii)  in a component of the polar motion or specific planetary ephemerides, on the other hand (Figures \ref{fig:02}-\ref{fig:10}). If local processes would have dominated the response of faults to these long oscillations, such phase lag would likely not have been consistent and maintained over time, and in particular, it would likely not be the same for the different periods extracted, due to local conditions (pore-fluid pressure, stress field, etc), as well as the heterogeneity in the internal properties (composition, porosity, permeability \etc) of rocks composing the Earth’s lithosphere. Similarly as for the results obtained for the analysis of the global volcanism (\shortciteNP{LeMouel2023}), the phase lag of  $\varphi$/4 which represents a shift related to temporal derivative of sinusoidal functions, was detected between the solid Earth and fluid envelopes on one hand, and the polar motion on another hand. These results are coherent with Laplace’s theory in which, the time shift expresses a causal chain of forces acting in response to orbital moments of the Jovian planets (\shortciteNP{Lopes2021,Lopes2022a} and \shortciteNP{LeMouel2023}, Figure 9).
	
	In addition to these observations, we would like to draw attention on the pseudo-cycles we extracted in this present study, they are featured by periods and waveforms that are more than compatible to even identical with those we had discovered in global volcanism (\shortciteNP{LeMouel2023}, Figure 11), again independently of the nature of the eruption and the volcano setting.  Moreover, a series of papers have demonstrated that probable similar forcing participate in the same way, in the triggering of volcanic eruptions, driven by variations in the sea level on long-term, (as shown in this study, \shortciteNP{Dumont2022,Dumont2023,Satow2021}), or on short-term, through a tidal modulation of the magmatic and hydrothermal fluids (\shortciteNP{DeLauro2013,Dumont2021,Petrosino2022}).
	
	Finally, we would like to note that the absence of the $\sim$11-year cycle in the SL@B (Table \ref{table:01}), is not a disqualifier: just because it does not occur in open water does not mean it is absent in confined environments, such as groundwaters. Recent works have shown that, at least since the end of the last millennium, aquifers and groundwater have been influenced by solar cycles (\eg \shortciteNP{Diodato2024,Fan2024}) and thus by the movement of the rotation pole. This is not surprising, given that the same $\sim$11-year component is one of the main factors in the variations of the Earth's magnetic field (and hence the dynamo, \shortciteNP{LeMouel2023b}), as seen in the LOD. The planetary orbital momenta in our solar system influence the rotational axis of the poles and, consequently, affect variations in the hydrosphere's levels, for instance, through LOD changes. While vertical groundwater movements in the subsurface are complex to track locally, on a global scale, they should generally follow sea-level variations. 

\paragraph{On the possible trigger mechanism}
	Earthquake triggering is as an intricate process reflecting the complexity of physico-chemical processes taking place in the Earth’s heterogeneous lithosphere (\shortciteNP{KeilisBorok1990}), \ie a hierarchical structure of blocks-and-faults extending from the scale of tectonic plates to that of grains of rock minerals. Earthquakes can be regarded as a critical transition of such non-linear complex systems. Similarly, the Earth’s network of faults may be seen as a metastable system whose instability reflects a stored energy originating in physical and/or chemical mechanisms like for instance, the Rehbinger effect \shortciteNP{Gabrielov1983}), non-linear filtration (\shortciteNP{Barenblatt1983}), "fingers" springing out at the front of a migrating fluid (\shortciteNP{Barenblatt1996}), sensitivity of dynamic friction to local environment (\shortciteNP{LomnitzAdler1991}), multiple fracturing, viscous flow, petrochemical transitions that tie up or release fluids, and much more. Actually, the instability of fault systems likely encompass both physical and chemical processes, not only physical ones as usually considered. The fact that many of the aforementioned mechanisms involve fluids is interesting as they represent a sort of interface between both processes, promoting chemical reactions resulting in variations of physical parameters at microscopic scale. Knowing that fluids are ubiquitous to our planet and are known to participate in rock failure by reducing effective stresses within the material (\shortciteNP{Gabrielov1983,Scholz2019,Wang2021}), we propose to consider our results from a new perspective, that of the fluid-rock interaction. 	
	
	 Recent studies based on hydrochemistry of groundwaters provide significant insights on how fluid-rock interactions could explain our observations. Here, we used the general term groundwaters to encompass all kind of subterranean waters present in the continental and oceanic crust (\shortciteNP{Fisher2005}). First, it is important to consider that not all minerals are in thermodynamic equilibrium with water under the conditions of groundwater stored in the Earth’s crust. Groundwaters can be thus seen as an evolutionary system allowing the thermodynamic balance between unstable and newly-formed yet stable mineral phases by promoting dissolution-precipitation processes (\shortciteNP{Zuddas2010}). This mainly leads to changes in water content, reflected in ion balance and composition (major elements, electrical conductivity), CO$_{2}$ release, redox conditions, pH levels, and variations in stable isotopic ratios such as $^{3}$He/$^{4}$He (\shortciteNP{Skelton2014,Skelton2019,Chiodini2020,Zuddas2024}). These changes, in turn, affect mineral surfaces and rock properties, especially porosity and permeability, which may affect the internal stability of rock.
	
	Actually, variations in water content have been identified within six months prior to the occurrence of earthquakes of magnitude $>$ 5 in basaltic material in North Iceland, results which were further used for testing forecasts based on last 10-year data (\shortciteNP{Skelton2019,Skelton2024}). These works have several implications, as they suggest that (1) destabilisation of rock structure may be critically detected within months prior to a large seismic event, (2) these fluid-rock interactions participate in rock weakening and microfracturing, (3) this kind of weathering process is not exclusively affecting carbonates, but silicates as well, as shown earlier and (4) this process takes place on relatively short timescales. In addition, the study by \shortciteN{Zuddas2024} implies that water-rock interactions which participate also to CO$_2$ release, which certainly mix/adds to that emitted from mantle depth (\shortciteNP{Chiodini2020}), may be forced mechanically by the oscillating movement of the groundwater forced by tides. Therefore, groundwaters appear as a dynamical system able to respond to sustained forcing and to act mechanically and chemically on the water-rock system. Moreover, as faults are more complex structures than just an oriented-surface plan, representing a discontinuity network in the rock matrice and thus preferred pathways for fluids, they provide expanded surface areas that can be submitted to such action. The dissolution-precipitation processes leading to progressive lost of consistency of the rock matrice and thus to its strength, may therefore be more easily prone to rupture when subjected to sustained oscillations even of lower amplitude as it may be the case of interannual to decadal variations, as those revealed with this study. One can note that the dissolution process is a complex and non-linear phenomenon, so we  cannot expect seismicity in a given region to respond directly to extraterrestrial forces. However, over long periods and from a global perspective, there is no reason to rule it out, and this is what our analysis indicates. This interaction would thus have two primary consequences: mineral dissolution, which weakens the ground, and the release of gases. Further works are necessary for testing and validating this hypothesis, requiring long and high resolution time-series of hydrochemistry and gas emissions such those performed in Italy by \shortciteN{Chiodini2020}  or in Iceland by  \shortciteN{Skelton2014},\shortciteN{Skelton2019}. 
	
\begin{figure}[H]
    \centering
    \includegraphics[width=\textwidth]{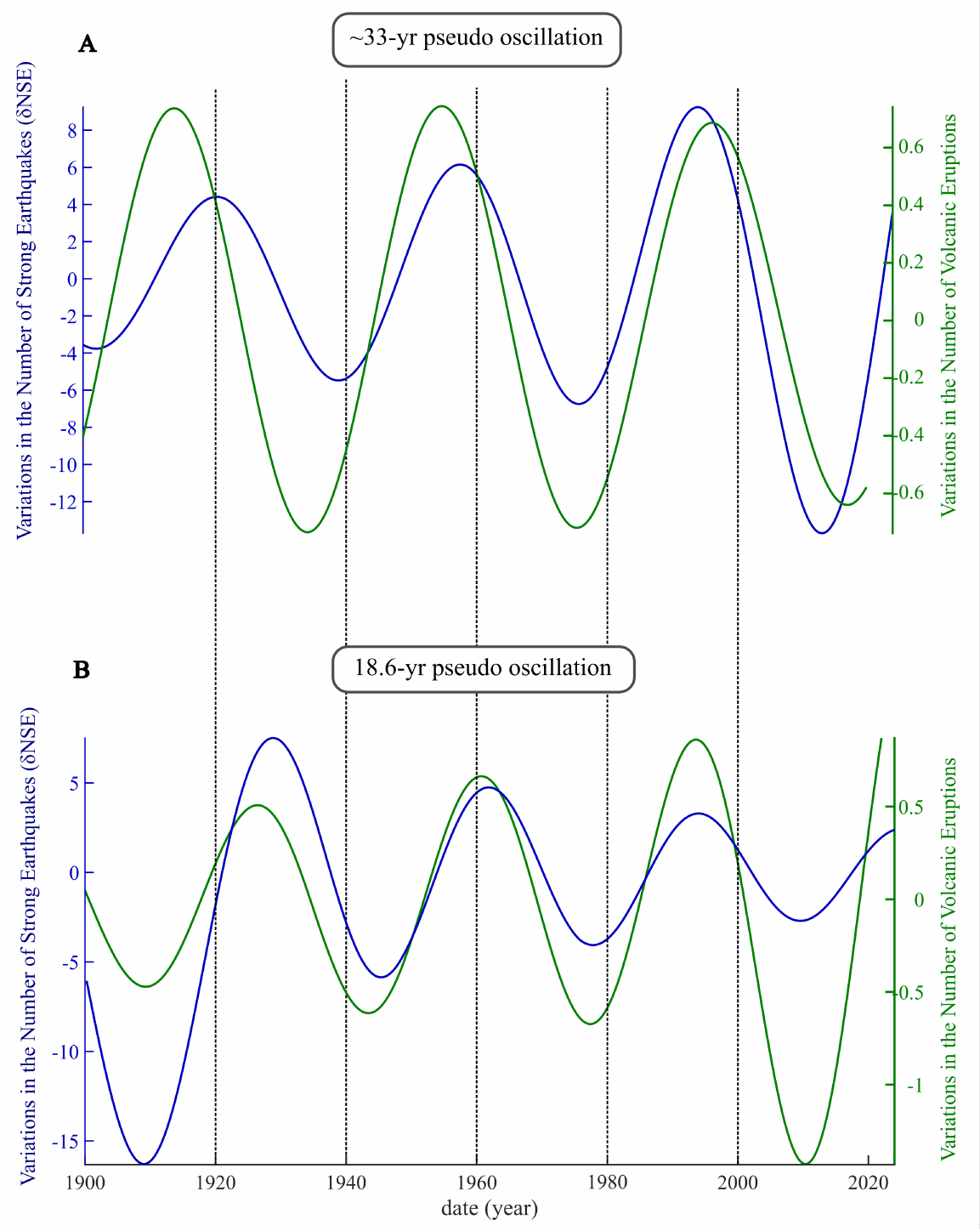}  
    \caption{ Comparison between the $\sim$33-year period (A) and the $\sim$19-year period (B) pseudo-cycles extracted from the global series of volcanic eruptions (green) (from Le Mouël et al. 2023) and the global series of strong earthquakes (blue, the present study)}
    \label{fig:11}
\end{figure}

	One final general comment. If the mechanism we suggest will be further confirmed, it should also play a role in volcanic eruptions. Volcanic environments, in addition to also host groundwater, are usually associated with hydrothermal systems which are recognized to be of critical importance in what deals with the system stability (\shortciteNP{Heap2021}). In Figure \ref{fig:11}, we superimpose the $\sim$33-year and the $\sim$19-year components extracted in the global volcanism (data from \shortciteNP{LeMouel2023}) and the corresponding components extracted in the NSE, revealing a strong in-phase correlation. 
	
	In summary, we aim at examining the temporal evolution of the global seismicity, focusing on the earthquakes of magnitude 6 or greater since 1900, in order to identify a common feature that might help us to deduce and propose a new general trigger mechanism. First, we present the dataset of seismic events (NSE) from January 1, 1900 to January 1, 2024, detailing its spatial distribution and general statistics. Next, we introduced two time-series related to tidal influences on Earth, specifically the LOD variations and sea level records from Brest, the latter being a representative gauge for general ocean levels, particularly in the Northern Hemisphere. These three series were decomposed \via SSA, revealing that they can be represented as the sum of seven quasi-periodic components related to the planetary commensurabilities of major Jovian planets, as well as the Sun, and the Moon. Altogether, these periods account for $\sim$88\% of the total energy in the seismic events signal, $\sim$89\% of the LOD variations, and 90\% of the sea level fluctuations at Brest (SL@B). They include, the annual oscillation,  the Quasi-Biennial Oscillation (QBO),  the $\sim$11-year, the $\sim$14-year, the $\sim$19-year, the $\sim$33-year, and finally, the $\sim$60-year pseudo-cycles, all known to manifest across various geophysical phenomena. We show that they are mainly in phase quadrature (\ie $\pi$/4) with planetary ephemerides. This constant, precise phase shift evokes Laplace’s equations in global mechanics, in which the hydrosphere is known to oscillate in quadrature with LOD variations. We discuss our results in the context of hydro-lithosphere interactions, focusing on water-rock interactions that reconcile both physical and chemical processes as well as the matter of scale.

\paragraph{Acknowledgment}
 This work was supported by the Portuguese Fundação para a Ciência e Tecnologia FCT I.P./MCTES through national funds (PIDDAC) –  UID/50019/2025, UIDP/50019/2020 (\url{https://doi.org/10.54499/UIDP/50019/2020}),LA/P/0068/2020 (\url{https://doi.org/10.54499/LA/P/0068/2020}), and FCT-through project RESTLESS (PTDC/CTA-GEF/6674/2020, \url{http://doi.org/10.54499/PTDC/CTA-GEF/6674/2020}).

\newpage
\bibliographystyle{fchicago}
\bibliography{seismicity_biblio.bib}
\newpage

\setcounter{section}{0}
\renewcommand{\thesection}{\Alph{section}}
\setcounter{figure}{0}
\renewcommand{\thefigure}{A\padzeroes[2]{\arabic{figure}}}
\section{\label{supp:A}Pseudo-cycles extracted from seismic events}
\begin{figure}[H]
    \centering
    \includegraphics[width=\textwidth]{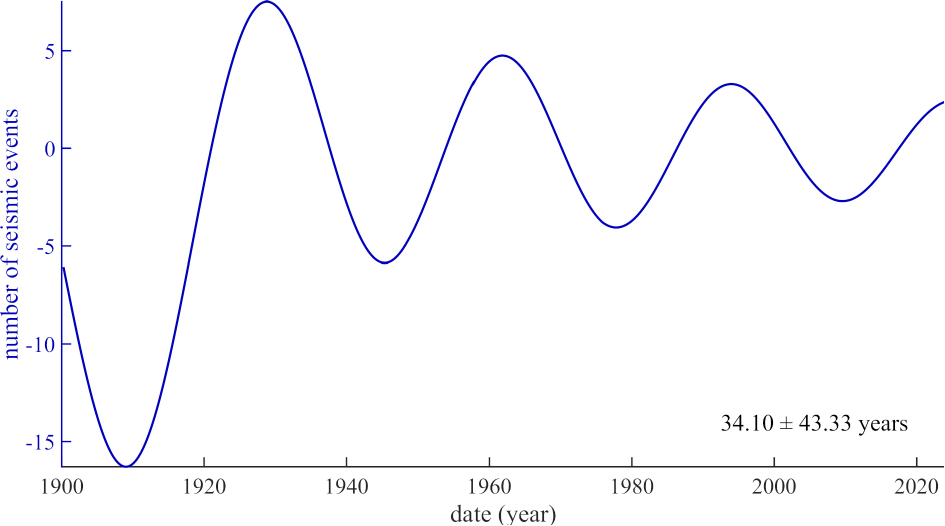}  
    \caption{The $\sim$ 33-years pseudo-cycles extracted from median earthquake series since 1900.}
    \label{fig:A01}
\end{figure}	
\begin{figure}[H]
    \centering
    \includegraphics[width=\textwidth]{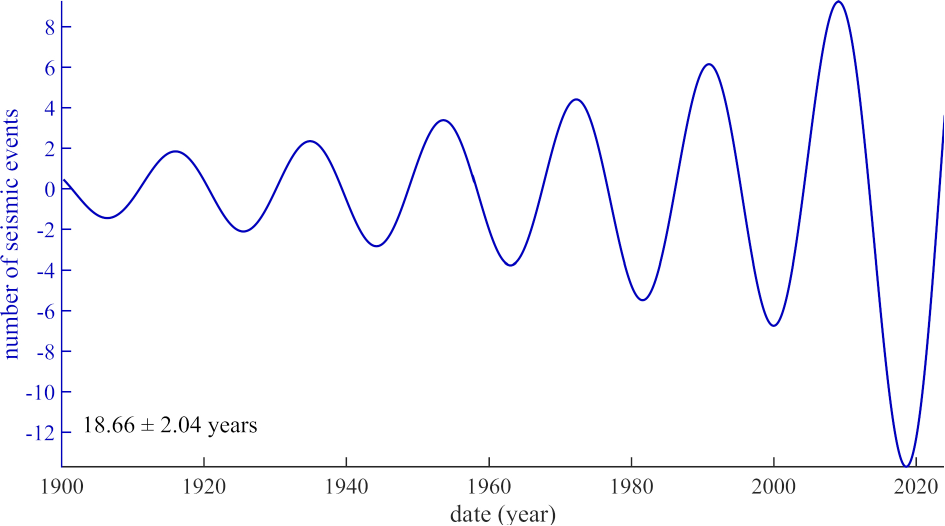}  
    \caption{The $\sim$18.6-years pseudo-cycles extracted from median earthquake series since 1900.}
    \label{fig:A02}
\end{figure}	
\begin{figure}[H]
    \centering
    \includegraphics[width=\textwidth]{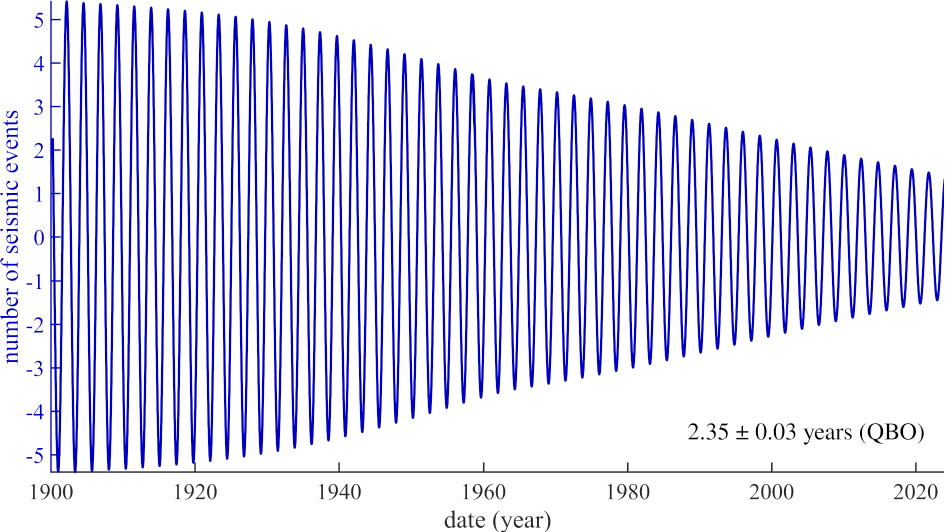}  
    \caption{The $\sim$2.3-years pseudo-cycles extracted from median earthquake series since 1900.}
    \label{fig:A03}
\end{figure}	
\begin{figure}[H]
    \centering
    \includegraphics[width=\textwidth]{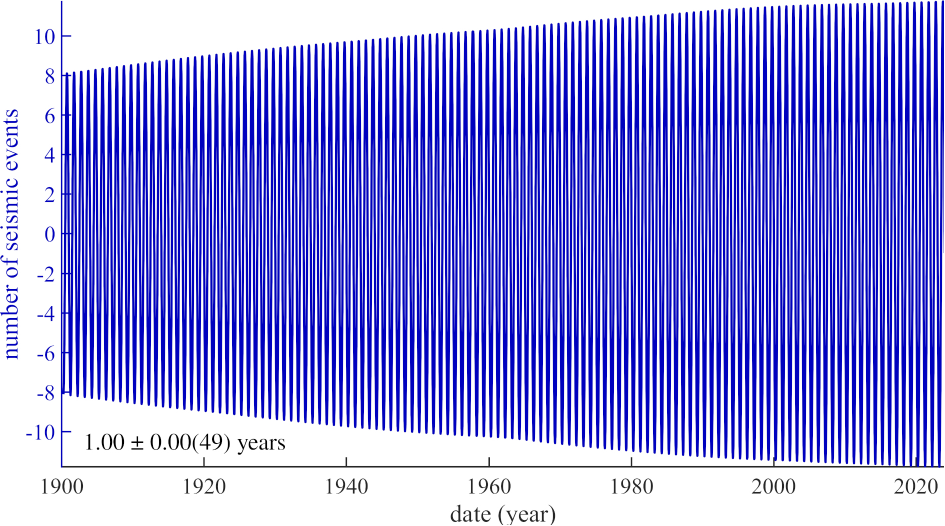}  
    \caption{The $\sim$1-year pseudo-cycles extracted from median earthquake series since 1900.}
    \label{fig:A04}
\end{figure}	
\begin{figure}[H]
    \centering
    \includegraphics[width=\textwidth]{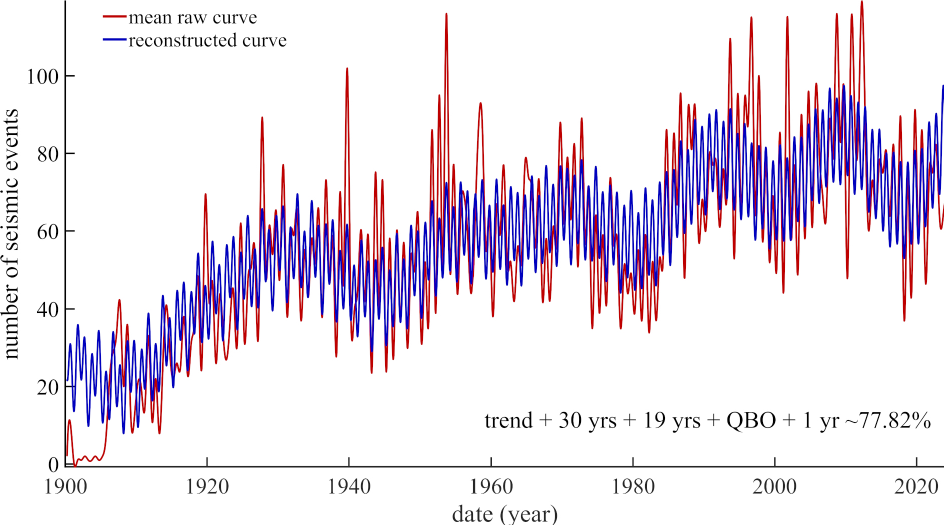}  
    \caption{Overlay of the raw signal (red curve), the reconstructed signal (blue curve) with the main pseudo-cycles detected and extracted.}
    \label{fig:A05}
\end{figure}	

\setcounter{figure}{0}
\renewcommand{\thefigure}{B\padzeroes[2]{\arabic{figure}}}
\section{\label{supp:B} Pseudo-cycles extracted from LOD}
\begin{figure}[H]
    \centering
    \includegraphics[width=\textwidth]{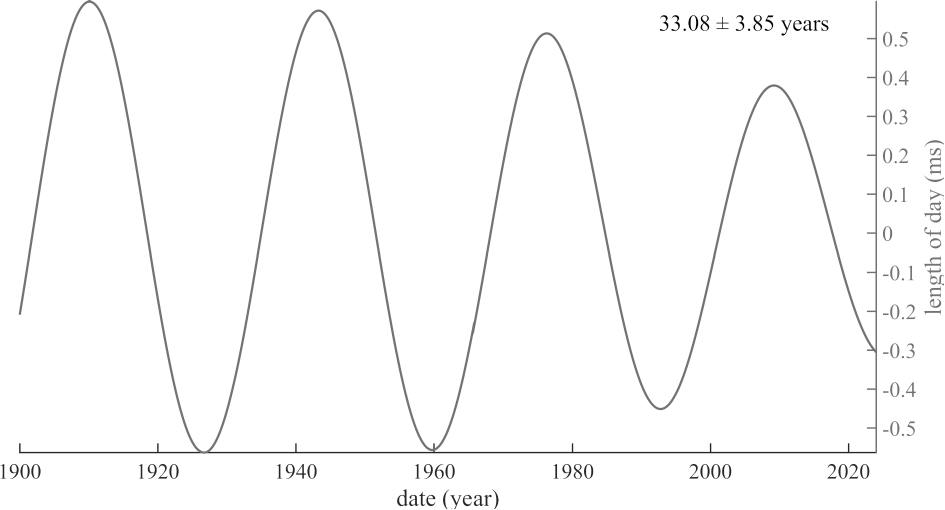}  
    \caption{The $\sim$33-years pseudo-cycles extracted from reconstructed  LOD series since 1900}
    \label{fig:B01}
\end{figure}	
\begin{figure}[H]
    \centering
    \includegraphics[width=\textwidth]{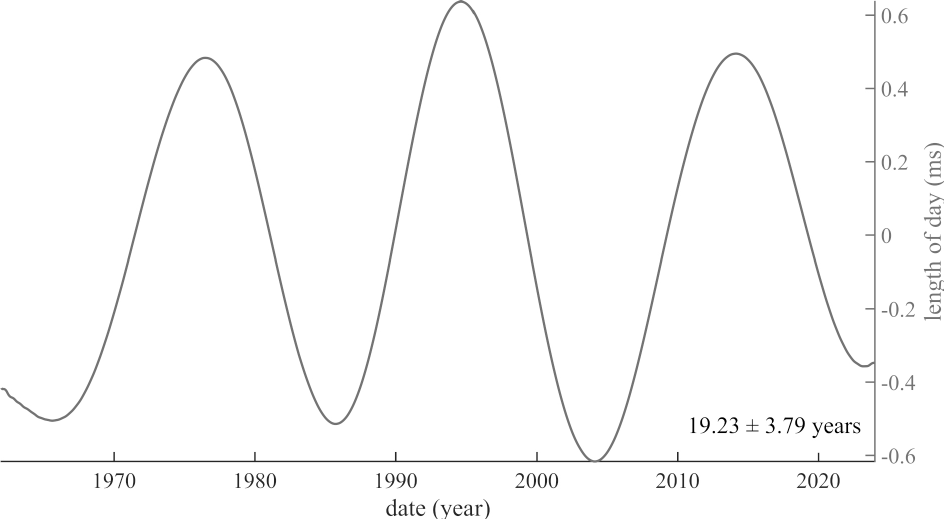}  
    \caption{The $\sim$18.6-years pseudo-cycles extracted from reconstructed  LOD series since 1900}
    \label{fig:B02}
\end{figure}	
\begin{figure}[H]
    \centering
    \includegraphics[width=\textwidth]{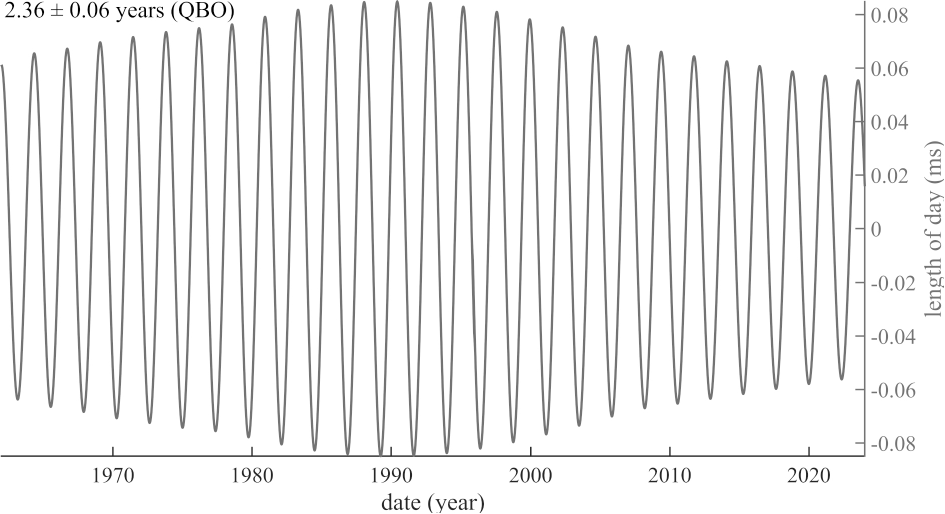}  
    \caption{The $\sim$2.3-years pseudo-cycles extracted from reconstructed  LOD series since 1900}
    \label{fig:B03}
\end{figure}	
\begin{figure}[H]
    \centering
    \includegraphics[width=\textwidth]{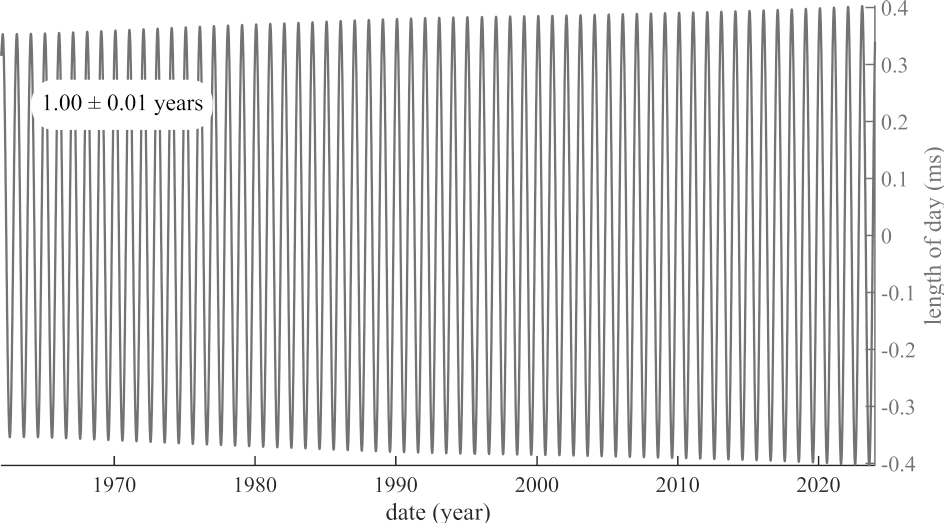}  
    \caption{The $\sim$1-year pseudo-cycles extracted from reconstructed  LOD series since 1900}
    \label{fig:B04}
\end{figure}	
\begin{figure}[H]
    \centering
    \includegraphics[width=\textwidth]{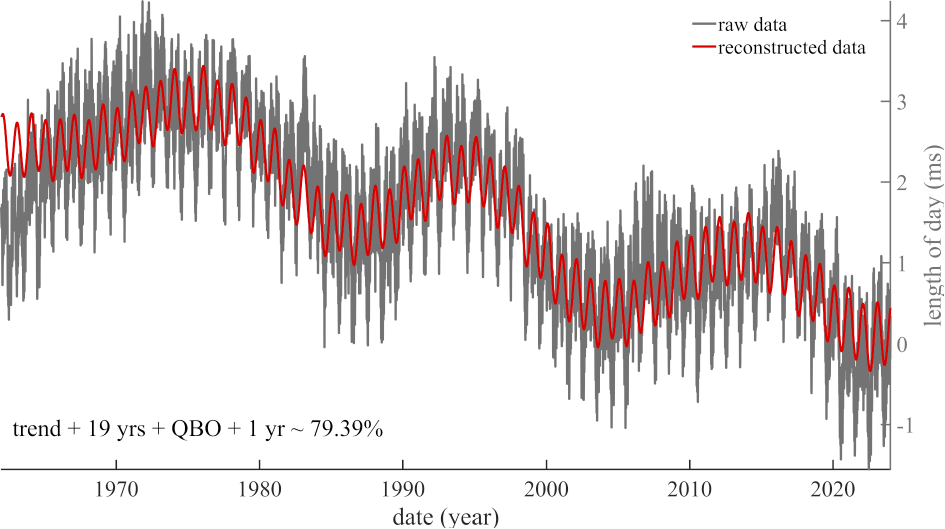}  
    \caption{Overlay of the raw signal (gray curve), the reconstructed signal (red curve) with the main pseudo-cycles detected and extracted.}
    \label{fig:B05}
\end{figure}

\setcounter{figure}{0}
\renewcommand{\thefigure}{C\padzeroes[2]{\arabic{figure}}}
\section*{\label{supp:C} Pseudo-cycles extracted from Brest tides gauge}
\begin{figure}[H]
    \centering
    \includegraphics[width=\textwidth]{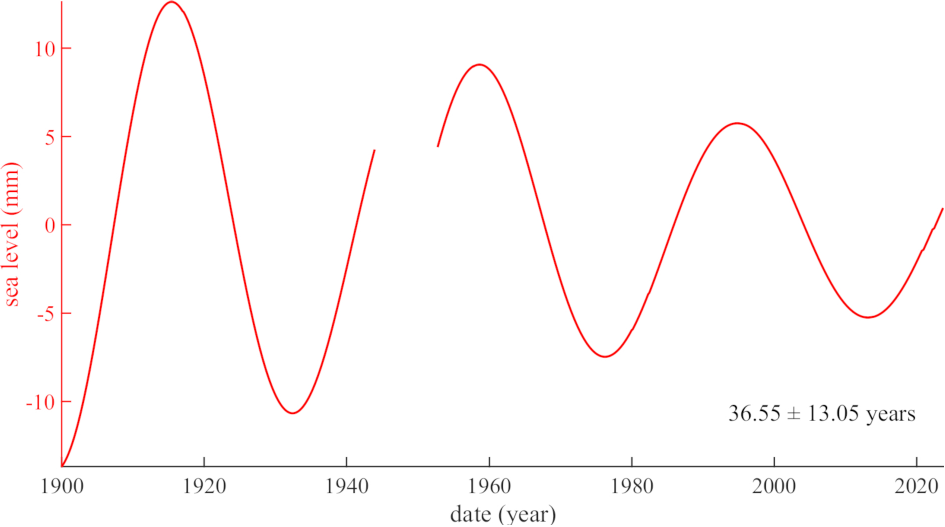}  
    \caption{The $\sim$33-years pseudo-cycles extracted from Brest tides gauge since 1900}
    \label{fig:C01}
\end{figure}	
\begin{figure}[H]
    \centering
    \includegraphics[width=\textwidth]{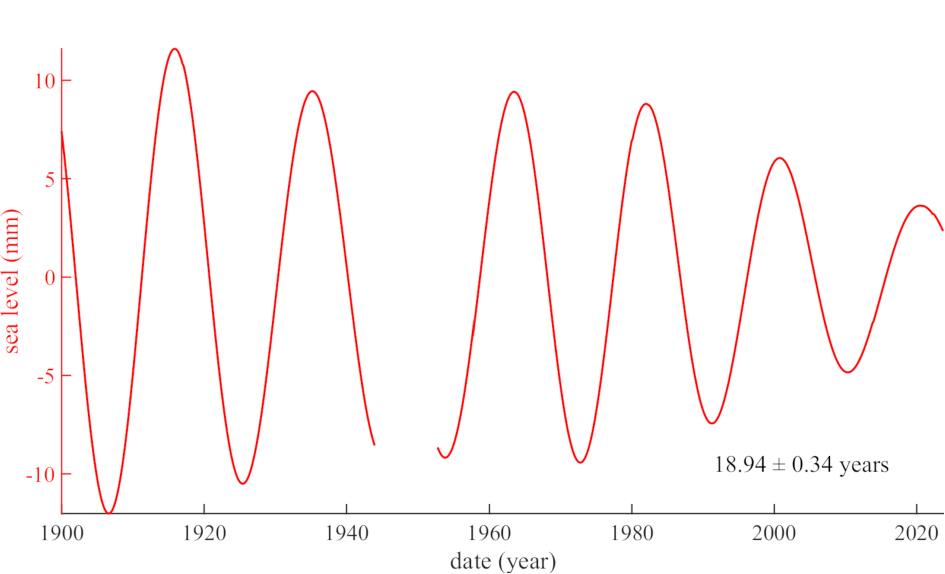}  
    \caption{The $\sim$18.6-years pseudo-cycles extracted from Brest tides gauge since 1900}
    \label{fig:C02}
\end{figure}	
\begin{figure}[H]
    \centering
    \includegraphics[width=\textwidth]{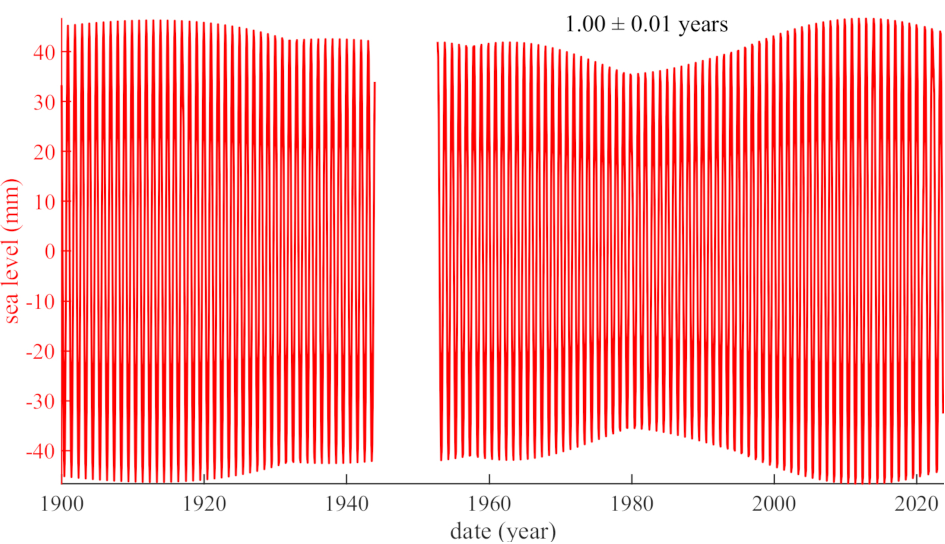}  
    \caption{The $\sim$1-year pseudo-cycles extracted from Brest tides gauge since 1900}
    \label{fig:C03}
\end{figure}	
\begin{figure}[H]
    \centering
    \includegraphics[width=\textwidth]{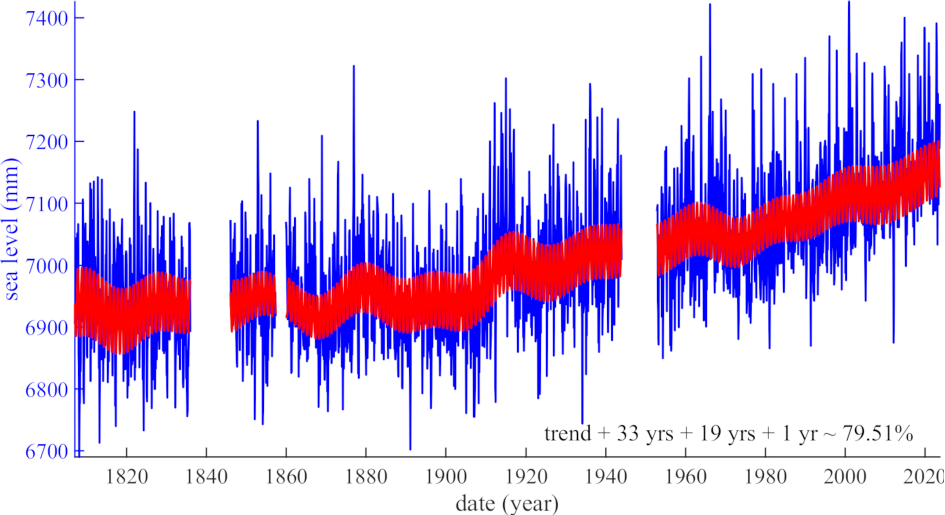}  
    \caption{Overlay of the raw signal (blue curve), the reconstructed signal (red curve) with the main pseudo-cycles detected and extracted.}
    \label{fig:C04}
\end{figure}	

\end{document}